\documentclass[10pt]{article}

\usepackage[superscript,sort]{cite}

\usepackage{ifthen,ifpdf}
\ifpdf
	\usepackage[pdftex]{hyperref}	\hypersetup{colorlinks=true,linkcolor=black,citecolor=black,urlcolor=blue,backref=page,bookmarks=true,breaklinks=true,plainpages=false }
	\usepackage[pdftex]{graphicx}
	\usepackage[usenames,pdftex]{color}
\else
	\usepackage[colorlinks=true,breaklinks=true,plainpages=false]{hyperref}
	\usepackage{graphicx}
	\usepackage[usenames]{color}
\fi

\usepackage{amsthm,amscd,amsxtra,amsfonts,amsmath,amssymb,multirow}
\usepackage{wrapfig}
\usepackage[footnotesize]{caption}
\usepackage[tiny,compact]{titlesec}
\usepackage[textwidth=0.8in,textsize=footnotesize]{todonotes}
\usepackage{algorithm,algorithmic,extarrows}

\begin{document}

\title{Persistent topology for  cryo-EM data analysis
}

\author{
Kelin Xia$^1$ and
Guo-Wei Wei$^{1,2,3}$ \footnote{ Address correspondences  to Guo-Wei Wei. E-mail:wei@math.msu.edu}\\
$^1$Department of Mathematics \\
Michigan State University, MI 48824, USA\\
$^2$Department of Electrical and Computer Engineering \\
Michigan State University, MI 48824, USA \\
$^3$Department of Biochemistry and Molecular Biology\\
Michigan State University, MI 48824, USA \\
}

\date{\today}
\maketitle

\begin{abstract}
In this work, we introduce persistent homology for the analysis of cryo-electron microscopy (cryo-EM) density maps. We identify the topological fingerprint or topological signature of noise, which is widespread in cryo-EM data.   For low signal to noise ratio  (SNR) volumetric data, intrinsic topological features of biomolecular structures are  indistinguishable from noise. To remove noise, we employ  geometric flows which are found to preserve the intrinsic topological fingerprints of cryo-EM structures and diminish the topological signature of noise. In particular, persistent homology enables us to visualize the gradual separation of the topological fingerprints of cryo-EM structures from those of noise during the denoising process,  which gives rise to a practical procedure for prescribing a noise threshold to extract cryo-EM structure information from noise contaminated data after certain iterations of the geometric flow equation. To further demonstrate the utility of persistent homology for cryo-EM data analysis, we consider a microtubule intermediate structure (EMD-1129).  Three helix models, an alpha-tubulin  monomer  model, an alpha- and beta-tubulin  model, and an alpha- and beta-tubulin dimer model,   are constructed to fit the cryo-EM data. The least square fitting leads to similarly high correlation coefficients, which indicates that structure determination via optimization is an ill-posed inverse problem.  However, these models have dramatically different topological fingerprints. Especially, linkages or connectivities that discriminate one model from another one,  play little role in the traditional density fitting or optimization, but are very sensitive and crucial to  topological fingerprints. The intrinsic topological features of the  microtubule data are identified after topological denoising. By a comparison of the topological fingerprints of the original data and those of three models, we found that the third model is topologically favored.  The present work offers  persistent homology based new strategies  for topological  denoising and for resolving ill-posed inverse problems.

\end{abstract}


\newpage

{\setcounter{tocdepth}{5} \tableofcontents}

\newpage

\section{Introduction}

The quantitative understanding of structure, function, dynamics and transport of biomolecules is a fundamental theme in contemporary life sciences. Geometric analysis and associated biophysical modeling have been the main workhorse in revealing the structure-function relationship of biomolecules and contribute enormously to the present understanding of biomolecular systems.  However, biology encompasses over more than twenty orders of magnitude in time scales from electron  transfer and ionization on the scale of femtoseconds to  organism life spanning over tens  of years, and over fifteen orders of magnitude in spatial scales from electrons and nuclei to  organisms.  The intriguing complexity and  extraordinarily large number of degrees of freedom of biological systems give rise to  formidable challenges to their quantitative description and theoretical prediction. Most biological processes, such as signal transduction, gene regulation,  DNA specification, transcription  and post transcriptional modification, are essentially  intractable for atomistic geometric analysis  and biophysical simulations, let alone {\it ab-initio} quantum mechanical descriptions.  Therefore, the complexity of biology and the need for its understanding offer an extraordinary opportunity for innovative theories, methodologies, algorithms and tools.

The study of subcellular structures, organelles and large multiprotein complexes  has become one of the major trends in structural biology. Currently, one of the most powerful tools for  the aforementioned systems is cryo-electron microscopy (cryo-EM), although other techniques, such as macromolecular X-ray crystallography, nuclear magnetic resonance (NMR), electron paramagnetic resonance (EPR),   multiangle light scattering, confocal laser-scanning microscopy, small angle scattering,  ultra fast laser spectroscopy, etc., are useful for structure determination in general   \cite{Nickell:2006,Robinson:2007,Leis:2009,Tocheva:2010, Volkmann:2010}. In cryo-EM experiments, samples  are  bombarded by electron beams at cryogenic temperatures to improve the signal to noise ratio (SNR). The  working principle is based on the projection (thin film) specimen scans collected from many different directions around one or two axes, and the Radon transform for the creation of three-dimensional (3D) images. One of major advantages of cryo-EM is that it allows the imaging of specimens in their native environment. Another major advantage is its  capability of providing 3D mapping of entire cellular proteomes together with their detailed interactions at nanometer or subnanometer resolution  \cite{Nickell:2006,Robinson:2007,Leis:2009,Tocheva:2010}.  The  resolution of cryo-EM maps has been improved dramatically in the past two decades, thanks to the technical advances in experimental hardware, noise reduction and image segmentation techniques. By further taking the advantage of symmetric averaging, many cryo-EM based virus structures have already achieved a resolution that can be interpreted in terms of atomic models. There have been a variety of  elegant methods  \cite{Abeysinghe:2008,Biswas:2012,TJu:2007,MBaker:2006,WSun:2009,YLu:2008} and software packages in cryo-EM structural determination  \cite{UCSFChimera:2004,UCSFTomography:2007,RAPTOR:2008,PyTom:2012,IMOD:1996,EM3D:1999,Rohl:2004}.

Most biological specimens are extremely radiation sensitive and can only sustain a limited electron dose of  illumination. As a result,  cryo-EM images are inevitably of low SNR and limited resolution  \cite{Volkmann:2010}.  In fact, the SNRs of  cryo-tomograms for subcellular structures, organelles and large multi-protein complexes are typically in the neighborhood of 0.01  \cite{Volkmann:2010}. To make the situation worse, the image contrast, which depends on the difference between electron scattering cross sections of cellular components, is also very low in most biological systems. Consequently,   cryo-EM maps often do not contain adequate information to offer unambiguous atomic-scale structural  reconstruction of biological specimens. Additional information obtained from other techniques, such as X-ray crystallography, NMR and computer simulation, is indispensable  to achieve  subnanometer resolutions. However, for cryo-EM data that do not have much  additional information obtained from other techniques, the determination of what proteins are involved can be a challenge, not to mention subnanometer structural resolution.

To improve the SNR and image contrast of cryo-EM data, a wide variety of denoising algorithms has been employed  \cite{Stoschek:1997,Frangakis:2001,Fernandez:2003,Fernandez:2009,Tomasi:1998,Jiang:2003,Pantelic:2006,vanderHeide:2007}. Standard techniques, such as bilateral filter  \cite{Tomasi:1998,Jiang:2003,Pantelic:2006} and iterative median filtering  \cite{vanderHeide:2007} have been utilized for noise reduction. Additionally,  wavelets and related techniques have also been developed for cryo-EM noise removing  \cite{Stoschek:1997}.  Moreover,  anisotropic diffusion  \cite{Frangakis:2001,Fernandez:2003} or Beltrami flow  \cite{Fernandez:2009} approach has been proposed for cryo-EM signal recovering.  However, cryo-EM data denoising  is far from adequate and remains a challenge due to the extremely low SNRs and other technical complications  \cite{Volkmann:2010,Tsai:2012,MPan:2011,Radaelli:2010}. For example, one of difficulties is how to distinguish signal from noise in cryo-EM data. As a result, one does not know when to stop or how to apply a threshold in an iterative noise removing process.  There is a pressing need for innovative mathematical approaches to further tackle this problem.

Recently, persistent homology has been advocated as a new approach for dealing with  big data sets  \cite{Fujishiro:2000, Carlsson:2009,Ghrist:2008, Doraiswamy:2013}. In general,  persistent homology characterizes the geometric features with persistent topological invariants by defining  a scale parameter relevant  to topological events. The essential difference between the persistent homology and traditional topological approaches is that traditional topological approaches describe the topology of a given object in  truly metric free or coordinate free representations, while persistent homology analyzes the persistence of the topological features  of a given object via a filtration process, which creates a family of  similar copies of the object at different spatial resolutions. Technically, a series of nested simplicial complexes is constructed from a  filtration process, which captures topological structures continuously over a range of spatial  scales. The involved topological features are measured by their persistent intervals. Persistent homology is able to embed geometric information to topological invariants so that ``birth"  and ``death" of  isolated components, circles, rings, loops, pockets, voids or cavities at all geometric scales  can be monitored by topological measurements.  The basic concept  of persistent homology was introduced by Frosini and Landi~ \cite{Frosini:1999} and  by Robins~ \cite{Robins:1999} in 1999 independently.  Edelsbrunner et al.~ \cite{Edelsbrunner:2002} introduced the first efficient computational algorithm, and Zomorodian and Carlsson  \cite{Zomorodian:2005} generalized the concept. A variety of elegant computational algorithms  has been proposed to track topological variations during the  filtration process  \cite{Bubenik:2007, edelsbrunner:2010,Dey:2008,Dey:2013,Mischaikow:2013}. Often, the persistent diagram can be visualized through barcodes  \cite{Ghrist:2008}, in which various horizontal line segments or bars are the homology generators lasted over filtration scales.    It has been applied to a variety of domains, including image analysis  \cite{Carlsson:2008,Pachauri:2011,Singh:2008,Bendich:2010}, image retrieval  \cite{Frosini:2013}, chaotic dynamics verification  \cite{Mischaikow:1999,Kaczynski:2004}, sensor network  \cite{Silva:2005}, complex network  \cite{LeeH:2012,Horak:2009}, data analysis  \cite{Carlsson:2009,Niyogi:2011,BeiWang:2011,Rieck:2012,XuLiu:2012}, computer vision  \cite{Singh:2008}, shape recognition  \cite{DiFabio:2011} and computational biology  \cite{Kasson:2007,Gameiro:2013,Dabaghian:2012}.

The concept of persistent homology has also been used for noise reduction. It is generally believed that short lifetime events (or bars) are of less importance and thus  regarded as ``noise'' while long lifetime ones are considered as ``topological signals''  \cite{Kloke:2009}, although this idea was challenged in a recent work  \cite{KLXia:2014c}. In topological data analysis, pre-processing algorithms are needed to efficiently remove these noise. Depending on the scale of a feature, a simple  approach is to pick up a portion of landmark points as a representative of topological data  \cite{DeSilva:2004}. The points can be chosen randomly, spatially evenly, or from extreme values. More generally, certain functions can be defined as a guidance for node selection to attenuate the noise effect, which is known as thresholding. Clustering algorithms with special kernel functions can also be employed to recover topological signal  \cite{Kloke:2009}.  All of these methods can be viewed as a process of data sampling  without losing the significant topological features. They rely heavily on the previous knowledge of the geometric or statistic information.  In contrast, topological simplification  \cite{Edelsbrunner:2002,Gunther:2014,Bauer:2010}, which is to  remove  the simplices and/or the topological attributes that do not survive certain threshold, focuses directly on the persistence of topological invariant. In contrast, Gaussian noise is known to generate a band of bars distributes over a wide range in the barcode representation \cite{Adler:2010}. Thank to the pairing algorithm, persistence of a homology group is measured through an interval represented by a simplex pair. If the associated topological invariant is regarded less important, simplices related to this simplex pair are reordered. This approach, combined with Morse theory, proves to be a useful tool for denoising  \cite{Gunther:2014,Bauer:2010}, as it can alters the data locally in the region defined as noise. Additionally, statistical analysis has been carried out to provide confidence sets for persistence diagram. However, persistent homology has not been utilized for cryo-EM data noise reduction, to our knowledge.

A large amount of experimental data for macroproteins and protein-protein complexes has been  obtained from  cryo-EM. To analyze these structural data, it is a  routine procedure to fit them with the available high-resolution crystal structures of their component proteins. This approach has been shown  to be efficient for analyzing many structures and has been integrated into many useful software packages  such as Chimera  \cite{Pettersen:2004}. However, this docking process is limited by data quality. For some low resolution data, which usually also suffer from low SNRs, there is enormous ambiguity in  structure fitting or optimization, i.e., a mathematically ill-posed inverse problem. Sometimes, high correlation coefficients can be attained simultaneously in many alternative structures, while none of them proves to be biologically meaningful. Basically, the fitting or optimization emphasizes more on capturing  ``bulk'' regions, which is reasonable as greater similarities in density distributions  imply higher possibility. However, little attention is paid to  certain small ``linkage'' regions, which play important roles in biological system especially in macroproteins and protein-protein complexes. Different linkage parts generate different connectivity, and thus directly influence biomolecular flexibility, rigidity, and even its functions. Since persistent homology is equally sensitive to both bulk regions and small linkage regions,  it is able to make a critical judgment on the model selection in structure determination, However, nothing  has been reported on persistent homology based solution to ill-posed inverse problems, to our knowledge.

Although persistent homology has been applied to a variety of fields, the successful use of persistent homology is mostly limited to characterization, identification and analysis (CIA).  Indeed, persistent homology has seldom employed for quantitative prediction. Recently, we have introduced molecular topological fingerprints (MTFs) based on persistent homology analysis of molecular topological invariants of biomolecules  \cite{KLXia:2014c}. We have utilized MTFs to reveal the topology-function relationship of macromolecules. It was found that protein flexibility and folding stability  are strongly correlated to protein topological connectivity, characterized by the persistence  of topological invariants   (i.e., accumulated bar lengths) \cite{KLXia:2014c}.  Most recently, we have employed persistent homology to analyze the structure and function of nano material, such as nanotubes and fullerenes. The MTFs are utilized to  quantitatively predict total curvature energies of fullerene isomers  \cite{KLXia:2014d}.

The overall objective of this work is to explore the utility of persistent homology for cryo-EM analysis.  First, we propose a topology based algorithm for cryo-EM noise reduction and clean-up.  We study the topological fingerprint or topological signature of noise and its relation to the  topological fingerprint  of cryo-EM structures. We note that the  histograms of topological invariants of the Gaussian random noise have  Gaussian distributions in the filtration parameter space.  Contrary to the common belief that short barcode bars correspond to noise,  it is found that there is an inverse relation between the SNR and the band widths of  topological invariants, i.e., the lower SNR,  the larger barcode band width is. Therefore, at a low SNR, noise can produce long persisting topological invariants or bars in the barcode presentation.   Moreover, for cryo-EM data of low SNRs, intrinsic topological features of the biomolecular structure are hidden in the persistent barcodes of noise and indistinguishable from noise contributions. To recover the topological  features of biomolecular structures, geometric flow equations are employed in the present work. It is interesting to note that topological features of biomolecular structures persist, while  the topological fingerprint of noise moves to the right during the geometric flow iterations. As such, ``signal'' and noise separate from each other during the geometric flow based denoising process and make it possible to prescribe a precise noise threshold for the  noise removal after certain iterations. We demonstrate the efficiency of our persistent homology controlled noise removal algorithm for both synthetic data and cryo-EM density maps.

Additionally,  we introduce persistent homology as a new strategy for resolving the ill-posed inverse problem in cryo-EM structure determination.  Although  the structure determination of microtubule data EMD 1129 is used as an example,  similar problems are widespread in other intermediate resolution  and low resolution cryo-EM data. As  EMD 1129 is contaminated by noise, a preprocess of denoising is carried out by using our persistent homology controlled geometric flow algorithm. A helix backbone is obtained for the microtubule intermediate structure. Based on  the assumption that the voxels with high electron density values are the centers of  tubulin proteins, we construct three   different microtubule models, namely a monomer model,  a two-monomer model, and a dimer model.  We have found that all three models give rise to essentially the same high correlation coefficients, i.e., 0.9601, 0.9607 and 0.9604, with the cryo-EM data. This ambiguity in structure fitting is very common with intermediate and low resolution data. Fortunately, after our topology based noise removal, the topology fingerprint of microtubule data is very unique, which is true for all cryo-EM data or data generated by using other molecular imaging modalities.  It is interesting to note that  although three models offer the same  correlation coefficients with the cryo-EM data, their topological fingerprints are dramatically different.  It is  found that the topological fingerprint of the microtubule   intermediate structure (EMD 1129) can be captured only when two conditions are simultaneously satisfied: first,  there must exist two different types of monomers, and additionally, two type of monomers from dimers.  Therefore, based on  topological fingerprint analysis, we can determine that  only the third model is a correct model for microtubule data EMD 1129.

The rest of this paper is organized as follows. The essential methods and algorithms for geometric and topological modelings of biomolecular data are presented in Section \ref{Methods}. Approaches for geometric modeling, which are necessary for topological analysis,  are briefly discussed.  Methods for persistent homology analysis are described in detail.  We illustrate the use of topological methods with both synthetic volumetric data and cryo-EM density maps. Their persistence of topological invariants is represented by barcodes. The geometric  interpretation of the topological features is given. Section \ref{Sec:PH_noise} is devoted to the persistent homology based noise removal. The characterization of Gaussian noise is carried out over a variety of SNRs to understand noise topological signature. Based on this understanding, we design a persistent homology monitored and controlled algorithm for noise removal, which is implemented via the geometric flow.  Persistent homology guided denoising is applied to the analysis of a supramolecular filamentous complex. In Section \ref{Sec:PH_microtubule},  we demonstrate topology-aided structure determination of  microtubule cryo-EM data.    Several aspects are considered including helix backbone evaluation, coarse-grained modeling  and topology-aided structure design and evaluation. We show that topology is able to resolve ill-posed inverse problem. This paper ends with a conclusion.

\section{Geometric  and topological modelings of biomolecular data }\label{Methods}

Persistent homology has been utilized to analyze biomolecular data, which are collected by different experimental means, such as macromolecular X-ray crystallography, NMR, EPR,  etc. Due to their different origins, these data may be available in different formats, which requires appropriate topological tools for their analysis. Additionally, their quality, i.e., resolution and SNR varies for case to case, and thus, a preprocessing may be required. Moreover, although biomolecular structures  are not a prerequisite for persistent homology analysis, the understanding of biomolecular structure, function and dynamics is crucial for the interpretation of  topological results.  As a consequence, appropriate geometric modeling  \cite{XFeng:2013b} is carried out in a close association with topological analysis. Furthermore, information from geometric and topological modelings is in turn, very valuable for data preprocessing and denoising. Finally, topological information is shown to be crucial for geometric modeling, structural determination and ill-posed inverse problems.

\subsection{Geometric  modeling of biomolecules}\label{Sec:geometric}
\begin{figure}
\begin{center}
\begin{tabular}{c}
\includegraphics[width=0.5\textwidth]{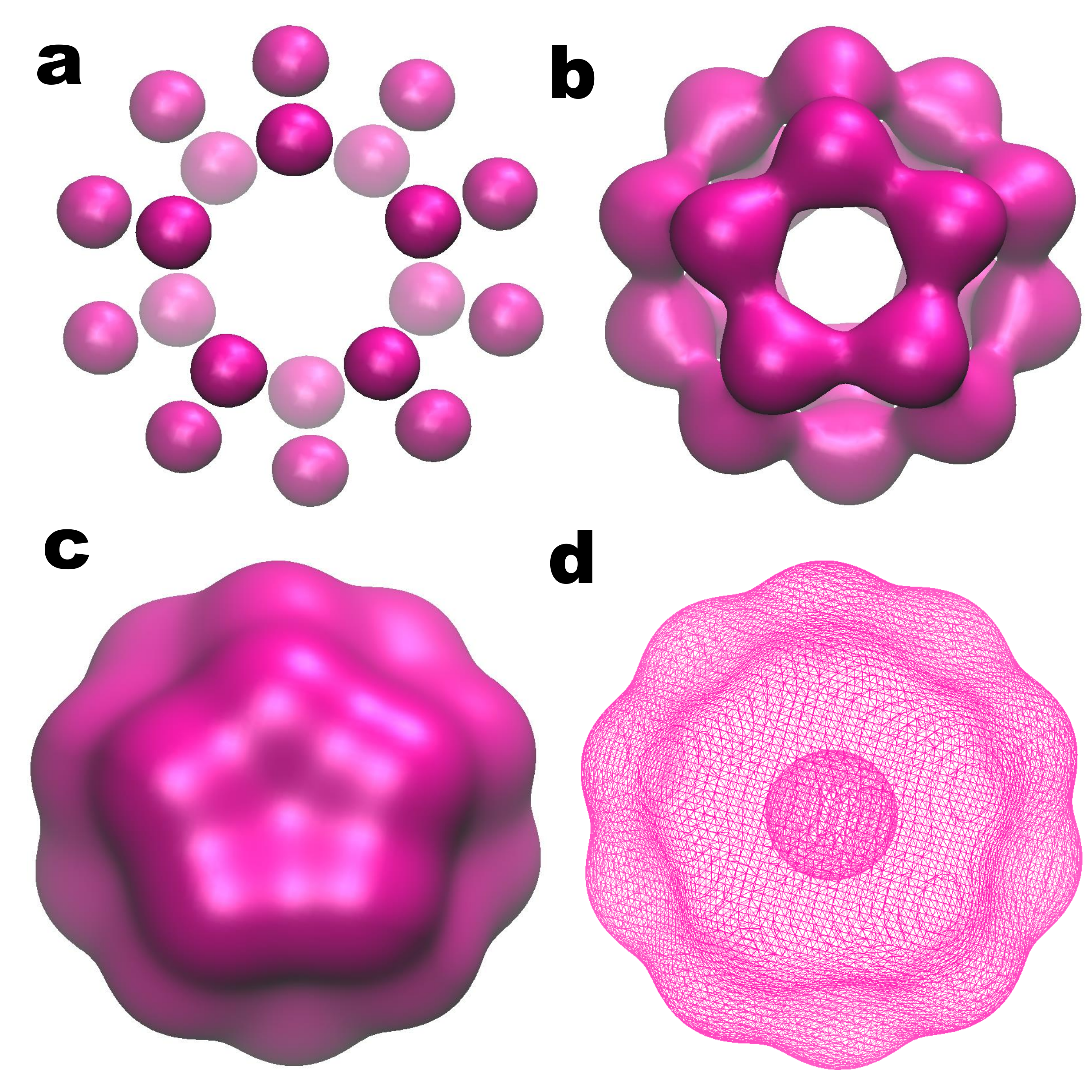}
\end{tabular}
\end{center}
\caption{A series of surfaces extracted from fullerene C$_{20}$ density function (\ref{eq:rigidity3}) with $w_j=1$, $\kappa=2$ and $\sigma_j=0.5$\AA. The isovalues for subfigures {\bf a}, {\bf b}  and  {\bf c} are 0.7, 0.6 and 0.4, respectively. Subfigure {\bf d} is wire-frame surface representation of subfigure {\bf c}. The Betti numbers can be directly obtained through identifying the numbers of connected components, circles or loops, and voids. In {\bf a}, $\beta_0$ is 20, $\beta_1$ and $\beta_2$ are 0; In {\bf b}, $\beta_0=1$, $\beta_1=11$ and $\beta_2=0$; In {\bf c} and {\bf d}, $\beta_0=1$, $\beta_1=0$ and $\beta_2=1$. Attention should be paid to $\beta_1$ for the structure in {\bf b}. The $\beta_1$ is not exactly equal to the total number  ($N_1$) of the circles or loops, and instead it is equal to $N_1-1$. This is due to the fact that the corresponding homology group has a basis with only $N_1-1$ elements. Therefore, one of elements can be expressed as the `` linear combination" of the rest. }
\label{fig:fullerene}
\end{figure}

Geometric modeling of biomolecules gives rise to their structural information, which is of paramount importance for biological understanding and structure-function relationship  \cite{XFeng:2013b,KLXia:2014a}.  Geometric modeling typically begins with experimental data.  There are two major repositories, namely,    \href{http://www.rcsb.org}{Protein Data Bank} (PDB)  and   \href{http://www.ebi.ac.uk/pdbe/emdb/}{Electron Microscopy Data Bank}  (EMDB), for storing biomolecular experimental data. The PDB lists detailed information of atomic   coordinates, occupancy and  Debye-Waller factor (or B-factor) of all the atoms in proteins, DNAs, RNAs and their complexes.  In the persistent homology terminology, PDB provides point cloud data for biomolecules. In contrast, the EMDB typically offers volumetric data, or density maps of large biomolecular systems, such as  multi-proteins, subcellular structures and  organelles   from cryo-EM  at the molecular level resolution.

Molecular structures and their visualization can be generated by using a variety of molecular models, such as the atom and bond model of molecules  \cite{Corey:1953},  the van der Waals surface, the solvent-excluded surface (SES) (also known as molecular surface (MS)) and the solvent-accessible surface, have been proposed  \cite{Lee:1971,Richards:1977}.  These models have been widely applied to the analysis of biomolecular structure, function, and interaction, such as ligand-receptor binding, protein specification, drug design, macromolecular assembly, protein-nucleic acid and protein-protein interactions, and enzymatic mechanism  \cite{Rocchia:2002}. Nevertheless, they admit geometric singularities, i.e., tips, cusps and self-intersecting surfaces which are  troublesome  in simulations   \cite{Connolly85,Eisenhaber:1993,Gogonea:1994,Sanner:1996} and {\it ad hoc} in physical foundation because electron density decays gradually at the molecular boundary  \cite{Wei:2009,ZhanChen:2010a}.

Recently, we have introduced the differential geometry theory of surfaces to address the above-mentioned  problems in biomolecular geometric modeling by curvature control PDEs    \cite{Wei:2005}, mean curvature flows  \cite{Bates:2006,Bates:2008} and potential driven geometric flows  \cite{Bates:2009}. The minimization principle was utilized for the biomolecular surface construction. We have further generalized these ideas to incorporate multiscale and multiphysical descriptions of biomolecules  \cite{Wei:2009,ZhanChen:2010a,Wei:2012,Wei:2013}. Our approaches have been adopted and/or generalized by many others   \cite{Xu:DSM:2006, Cheng:2007e, SZhao:2011a,SZhao:2014a}.

Most recently, we have proposed flexibility and rigidity index (FRI) for flexibility analysis and B-factor prediction of proteins and other biomolecules  \cite{KLXia:2013d,Opron:2014}.  In FRI, protein topological connectivity is measured by rigidity index and flexibility index. In particular, the rigidity index represents protein density profile. Consider a protein with $N$ atoms. Their locations are represented by $\{ {\bf r}_{j}| {\bf r}_{j}\in \mathbb{R}^{3}, j=1,2,\cdots, N\}$. We denote $  \|{\bf r}_i-{\bf r}_j\|$ the Euclidean space distance between  $i$th   atom  and the $j$th  atom. We   define a position (${\bf r}$)  dependent rigidity or density function  \cite{KLXia:2013d,Opron:2014}
\begin{eqnarray}\label{eq:rigidity3}
 \mu({\bf r}) & = & \sum_{j=1}^N w_{j}({\bf r}_j) \Phi( \|{\bf r} - {\bf  r}_j \|;\sigma_{j} ),
 \end{eqnarray}
where $w_{j}({\bf r})$ is an atom type dependent weight function and $\sigma_{j}$ is an atomic type dependent scale parameter which is proportional to the atomic radius.
 Here $ \Phi( \|{\bf r} - {\bf r}_j \|;\sigma_{j})$ is a  correlation kernel, which is, in general,  a  real-valued monotonically decreasing function satisfying
\begin{eqnarray}\label{eq:couple_matrix1-1}
\Phi( \|{\bf r} - {\bf r}_j \|;\sigma_{j})&=&1 \quad {\rm as }\quad  \|{\bf r} - {\bf r}_j \| \rightarrow 0\\\
\Phi( \|{\bf r} - {\bf r}_j \|;\sigma_{j})&=&0 \quad {\rm as }\quad  \|{\bf r} - {\bf r}_j \| \rightarrow\infty.
\end{eqnarray}
Although  Delta sequences of the positive type discussed in an earlier work  \cite{GWei:2000} are all good choices,   generalized exponential  functions
\begin{eqnarray}\label{eq:couple_matrix1}
\Phi(\|{\bf r} - {\bf r}_j \|;\sigma_{j}) =    e^{-\left(\|{\bf r} - {\bf r}_j \|/\sigma_{j}\right)^\kappa},    \quad \kappa >0
\end{eqnarray}
and  generalized Lorentz functions
\begin{eqnarray}\label{eq:couple_matrix2}
 \Phi(\|{\bf r} - {\bf r}_j \|;\sigma_{i}) = \frac{1}{1+ \left( \|{\bf r} - {\bf r}_j \|/\sigma_{j}\right)^{\upsilon}},  \quad  \upsilon >0.
 \end{eqnarray}
have been commonly used in our recent work    \cite{KLXia:2013d,Opron:2014}. We refer these general classes of volumetric surface definition as rigidity surfaces and density profiles.  We use a surface extraction procedure, such as marching cubes, to extract a Lagrangian surface from their volumetric data, or the Eulerian representation of surface density profiles. Obviously, when $\kappa=2$ in Eq. (\ref{eq:couple_matrix1}), Eq. (\ref{eq:rigidity3}) gives rise to a representation of Gaussian surfaces, which have many formulations  \cite{YZhang:2006, ZYu:2008,QZheng:2012,KLXia:2013,KLXia:2013f,KLXia:2013d,KLXia:2014a }. In general, Gaussian surfaces are quite smooth and free of geometric singularity. The generation of Gaussian surfaces can be very fast  and readily available in the Cartesian representation  \cite{QZheng:2012}. Other  geometric modeling approaches include curvature analysis and symmetry analysis.  Mean curvature and Gauss curvature can be estimated in both Lagrangian  representations  \cite{XFeng:2013b}and Eulerian representation  \cite{KLXia:2014a}. Maximal and minimal principle curvatures can be utilized for drug binding site prediction   \cite{KLXia:2014a}. Symmetric analysis is frequently employed in biophysical modeling \cite{QZheng:2012}. Utilizing symmetry leads to the reduction of number of genes, which is very common in  viral complexes. Many protein complexes, such as microtubules,  are  very symmetric as well.

Figure \ref{fig:fullerene} illustrates surfaces extracted from  density function Eq. (\ref{eq:rigidity3}) with   $w_j=1$, $\kappa=2$ and $\sigma_j=0.5$\AA. In this work, we  use density function (\ref{eq:rigidity3}) as a mathematical model for cryo-EM density maps. A series of surfaces is plotted in Fig.  \ref{fig:fullerene} to demonstrate  some typical structures in the filtration procedure  for fullerene C$_{20}$ density function. The isovalues for Figs. \ref{fig:fullerene} {\bf a}, {\bf b} and {\bf c} are 0.7, 0.6 and 0.4, respectively. Figure \ref{fig:fullerene} {\bf d} is a wire-frame surface representation of Fig. \ref{fig:fullerene} {\bf c}.

\subsection{Topological  modeling of biomolecules}\label{Sec:barcode}


\begin{figure}
\begin{center}
\begin{tabular}{c}
\includegraphics[width=0.5\textwidth]{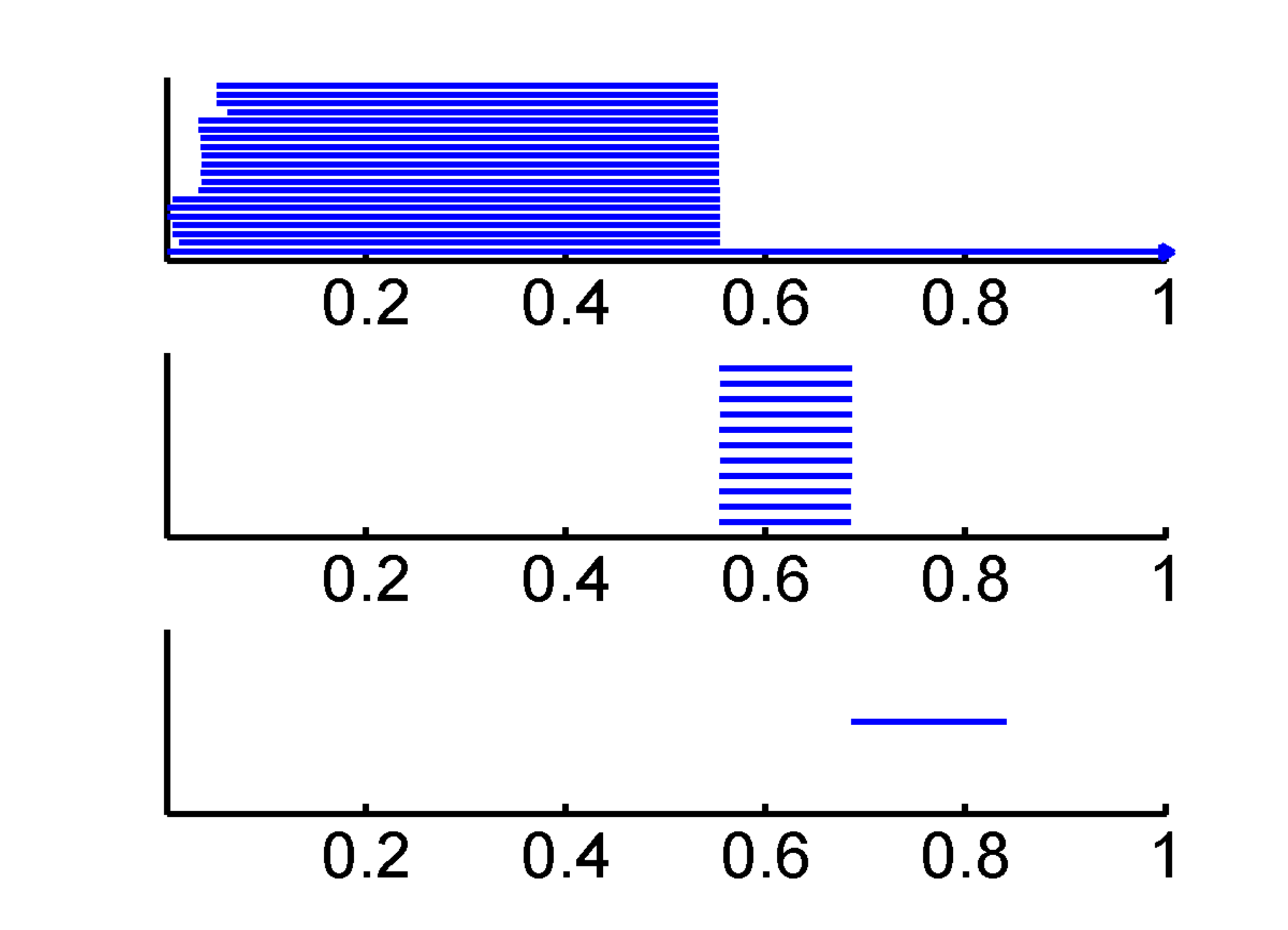}
\end{tabular}
\end{center}
\caption{The intrinsic topological fingerprints of fullerene C$_{20}$.  The top, middle, and bottom panels are for the barcodes of $\beta_0$,  $\beta_1$ and $\beta_2$, respectively.
The filtration process is based on the decrement of the density  isovalues. The filtration barcodes demonstrate pentagonal structures and central void. It should be notice that  $\beta_0$ bars do not emerges simultaneously, which is due to the Cartesian representation  of  data. }
\label{fig:Fulleren-Barcodes}
\end{figure}

Persistent homology theory and algorithm can be found in the literature  \cite{Edelsbrunner:2002,Zomorodian:2005,Bubenik:2007, Ghrist:2008, edelsbrunner:2010, Dey:2008, Mischaikow:2013} as well as our papers  \cite{KLXia:2014c, KLXia:2014d}. In this work, we  focus on electron density maps of macro-protein or protein-protein complexes available as volumetric data deposited in the EMDB. Unlike  point cloud data which are commonly studied with simplicial complex using Javaplex  \cite{javaPlex}, volumetric data are usually analyzed by the discrete Morse theory. For all the density map based volumetric data used in this work, we use the same filtration process that is built based on the decrement of the electron density value. More specifically, cryo-EM  density  maps or   density functions generated from Eq. (\ref{eq:rigidity3}) is used as the filtration parameter. The filtration process goes from the highest density isovalue to the lowest one. After the filtration, the data are analyzed with Perseus   \cite{Perseus}. We first consider a benchmark test to illustrate the persistent homology analysis of density function (\ref{eq:rigidity3}).

\paragraph{Topological persistence of C$_{20}$ }
 The Betti numbers can be directly obtained through identifying the number of connected components, circles or loops, and voids or holes. For example, in Fig.  \ref{fig:fullerene} {\bf a}, $\beta_0$ is 20, $\beta_1$ and $\beta_2$ are 0. In Fig.  \ref{fig:fullerene} {\bf b}, $\beta_0=1$, $\beta_1=11$ and $\beta_2=0$. In Fig.  \ref{fig:fullerene} {\bf c} and {\bf d}, $\beta_0=1$, $\beta_1=0$ and $\beta_2=1$. Attention should be paid to $\beta_1$ for the structure in Fig.  \ref{fig:fullerene} {\bf b}. The $\beta_1$ is not exactly equal to the total number  ($N_1$) of the circles or loops, instead it equals to $N_1-1$. This is due to the reason that the corresponding homology group has a basis with only $N_1-1$ independent elements. Roughly speaking, one of the circles can be expressed as the ``linear combination'' of the rest.


Barcodes provide a systematic representation of topological persistence   \cite{Ghrist:2008}. Figure  \ref{fig:Fulleren-Barcodes} shows the intrinsic topological patterns of fullerene C$_{20}$. Just the same as we counted above, there are 11 $\beta_1$ bars and only one $\beta_1$ bar. It should be noticed that   $\beta_0$ bars do not emerge simultaneously,  which is due to the discretization effect, namely, the density function is discretized with only a finite resolution.

\begin{figure}
\begin{center}
\begin{tabular}{c}
\includegraphics[width=0.5\textwidth]{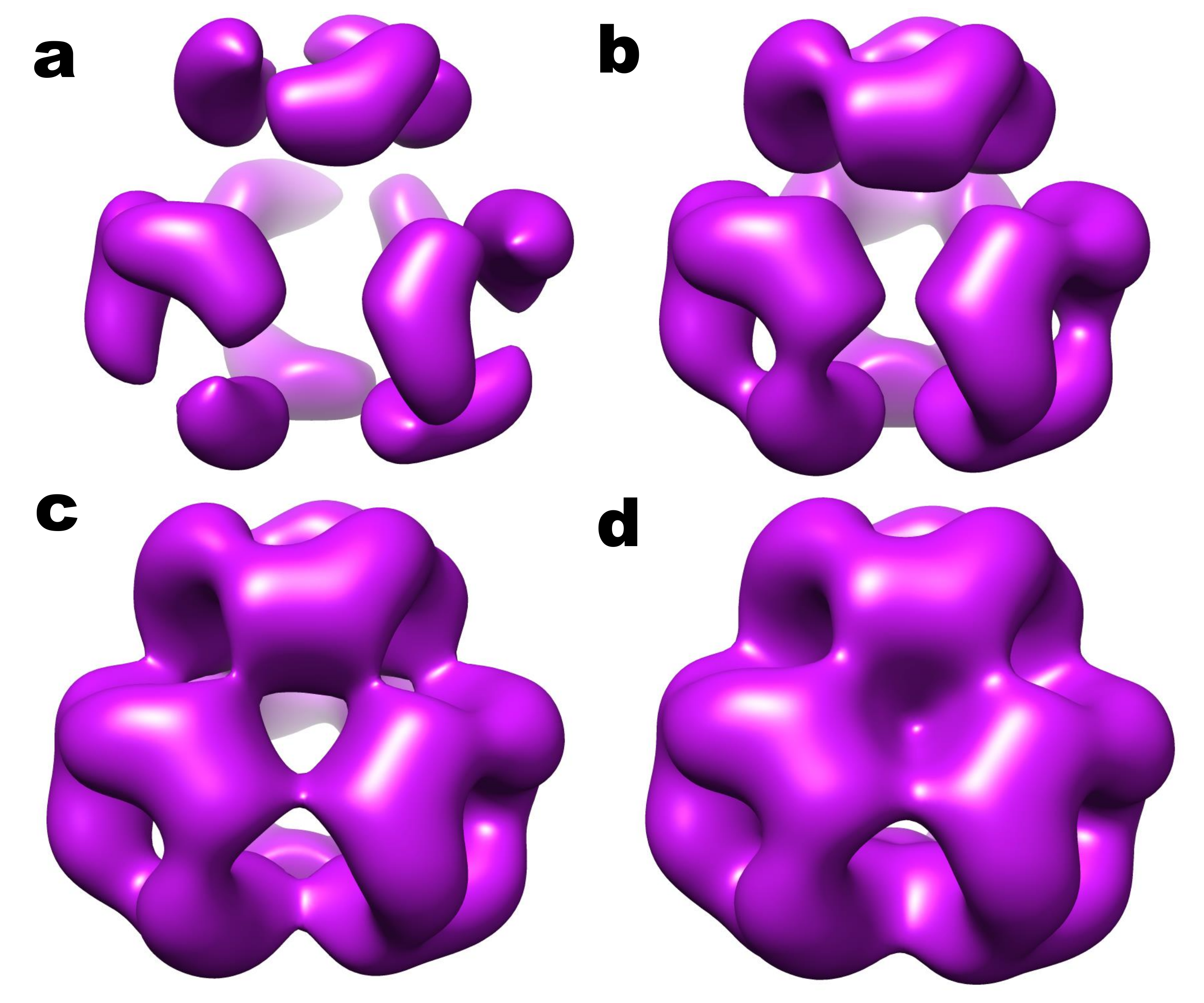}
\end{tabular}
\end{center}
\caption{Surfaces extracted with different isovalues for EMD 1776. The isovalues for subfigures {\bf a}, {\bf b}, {\bf c} and {\bf d} are 0.150, 0.100, 0.081 and 0.050 respectively. The Betti numbers can be directly obtained through identifying the number of connected components, circles or loops, and voids or holes. In {\bf a}, $\beta_0$ is 12, $\beta_1$ and $\beta_2$ are 0; In {\bf b}, $\beta_0=4$, $\beta_1=4$ and $\beta_2=0$; In {\bf c}, $\beta_0=1$, $\beta_1=13$ and $\beta_2=0$; In {\bf d}, $\beta_0=9$, $\beta_1=13$ and $\beta_2=0$. In {\bf c} and {\bf d}, the $\beta_1$ is one count fewer than  the total number  ($N_1$) of the circles or loops, because the corresponding homology group has a basis with only $N_1-1$ elements.}
\label{fig:EMD1776}
\end{figure}

\begin{figure}
\begin{center}
\begin{tabular}{c}
\includegraphics[width=0.5\textwidth]{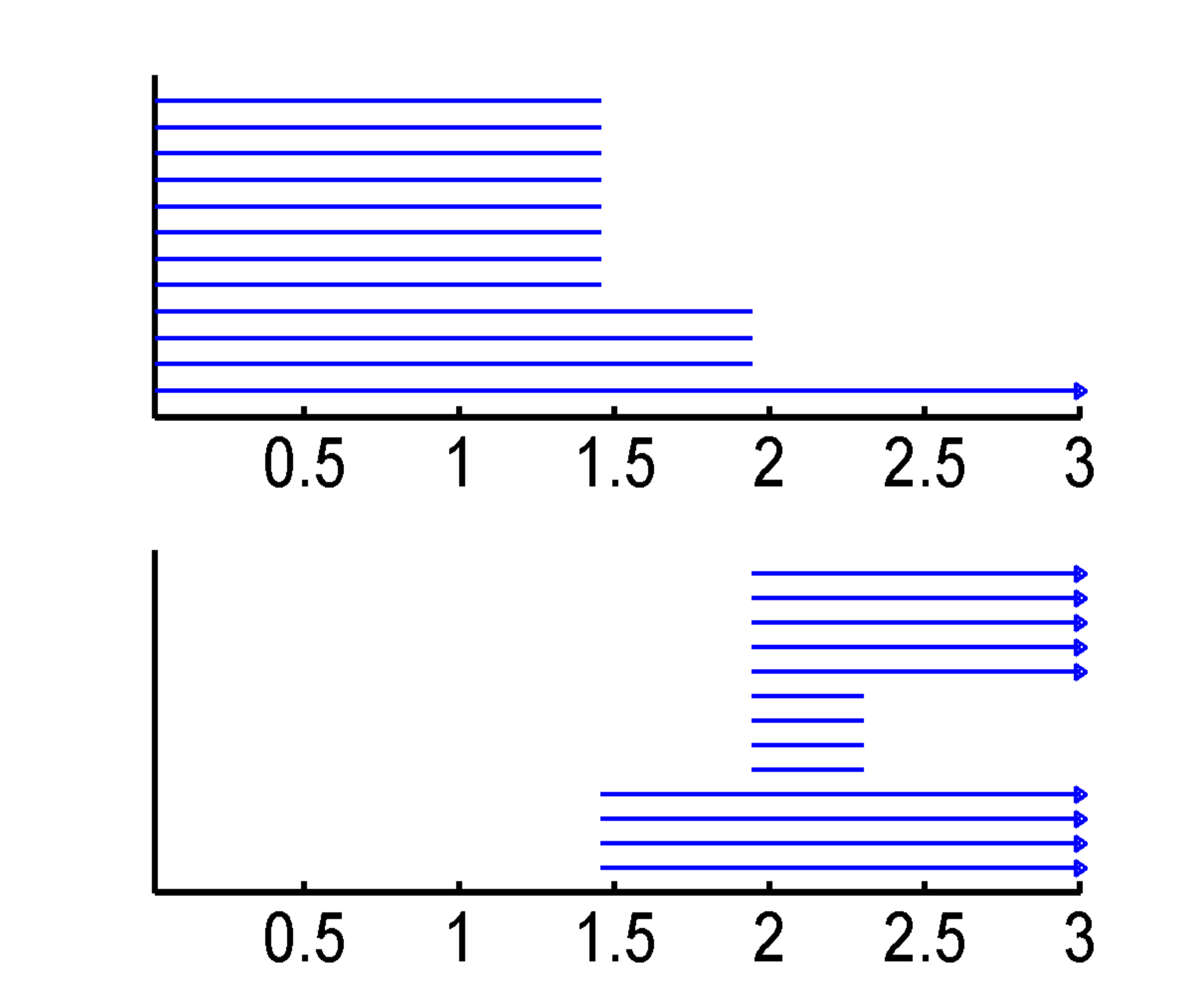}
\end{tabular}
\end{center}
\caption{The intrinsic topological patterns of the EMD 1776 structure.  The top  and bottom panels are for the barcodes of $\beta_0$ and $\beta_1$, respectively. The filtration process is based on the decrement of  density isovalues. The filtration goes from the largest isovalue (0.28) to isovalue threshold  (0.03). }
\label{fig:EMD1776-Barcodes}
\end{figure}

\paragraph{Topological persistence of  EMD 1776 }
After demonstrating the persistent homology analysis for  the density function (\ref{eq:rigidity3}) of a known structure (C$_{20}$), we further consider realistic cryo-EM data, EMD 1776, which is for  eye lens chaperone $\alpha$-crystallin assemblies. Figure  \ref{fig:EMD1776} depicts the surfaces extracted from different isovalues for EMD 1776. The isovalues for Figs.  \ref{fig:EMD1776} {\bf a}, {\bf b}, {\bf c} and {\bf d} are 0.150, 0.100, 0.081 and 0.050 respectively. Similarly,  Betti numbers can be directly obtained through counting the numbers of connected components, circles, and voids. In {\bf a}, $\beta_0$ is 12, $\beta_1$ and $\beta_2$ are 0. In Fig. \ref{fig:EMD1776}  {\bf b}, $\beta_0=4$, $\beta_1=4$ and $\beta_2=0$. In Fig. \ref{fig:EMD1776} {\bf c}, $\beta_0=1$, $\beta_1=13$ and $\beta_2=0$. Also in Fig. \ref{fig:EMD1776} {\bf d}, $\beta_0=9$, $\beta_1=13$ and $\beta_2=0$. As discussed above, in Figs. \ref{fig:EMD1776} {\bf c} and {\bf d}, the $\beta_1$ value is $N_1-1$, rather than  the total number  ($N_1$) of the circles, due to $N_1-1$ independent elements in the corresponding homology group.  The barcode representation is demonstrated in Fig.  \ref{fig:EMD1776-Barcodes}, which is consistent with our analysis.

It should be noticed that we only  consider the regions with density values larger than 0.03, namely, the filtration goes from the largest value (0.28) to a threshold value (0.03).  For density values smaller than 0.03, data suffer from lower SNRs as discussed in Section \ref{Sec:Topological_fingerprint}. Denoising techniques are  indispensable  for extracting more information from low isovalues. In the next section, we apply persistent homology to  noise reduction and topological feature identification.

\section{Persistent homology based  noise reduction}\label{Sec:PH_noise}

In this section, we  present persistent homology based cryo-EM data noise reduction, which is a crucial process  in cryo-EM analysis.  Protein data in EMDB are mostly obtained by cryo-EM.  As discussed earlier,  the  cryo-EM data suffer from  low resolution and low SNR. Therefore a  denoising process   is always a necessity before carrying out  geometric modeling and/or topological analysis.  We focus on noise reduction based on persistent homology analysis. Specifically, we employ persistent homology to discriminate signal from noise and  utilize this information for denoising thresholding.  In our benchmark test, we assume a known object is contaminated with Gaussian noise. Geometric flows proposed in our earlier work  \cite{Wei:1999,Bates:2009} are used for noise reduction of realistic cryo-EM data.

\subsection{Topological fingerprints of Gaussian noise}\label{Sec:HighOrderGF}

We first analyze the topological fingerprint or topological signature of noise. We use Gaussian noise as an example for the present study. Other noise can be analyzed in a similar manner.  The Gaussian white noise is generated by randomly selecting values from a normal distribution,
\begin{eqnarray}
n(t)=\frac{A_n}{ \sqrt{2\pi}\sigma_{n}}e^{-\frac{ (t-\mu_{n})^2}{2\sigma_{n}^2}},
\end{eqnarray}
where $A_n$, $\mu_{n}$ and $\sigma_{n}$ are the amplitude, mean value and standard deviation of the noise, respectively.  We denote  $\mu_{s}$ as the mean value of signal. Then the degree of noise contamination can be described by SNR, ${\rm SNR}=\mu_{s} / \sigma_{n}$. Please note that the present definition of SNR is by no means unique. Based on the physical properties of the signal, SNR can be defined in terms of average power, amplitude, variance of the signal, and so on. In our discussion, the signal information represented in volumetric data can be easily analyzed. We generate the noise data with specified SNR by adding suitable amplitude of Gaussian white noise. Stated differently, the noise contaminated data are generated by adding different levels of Gaussian white noise to the original data, i.e., the density function or density map. We then investigate  their corresponding persistent barcode patterns.

\begin{figure}
\begin{center}
\begin{tabular}{c}
\includegraphics[width=0.9\textwidth]{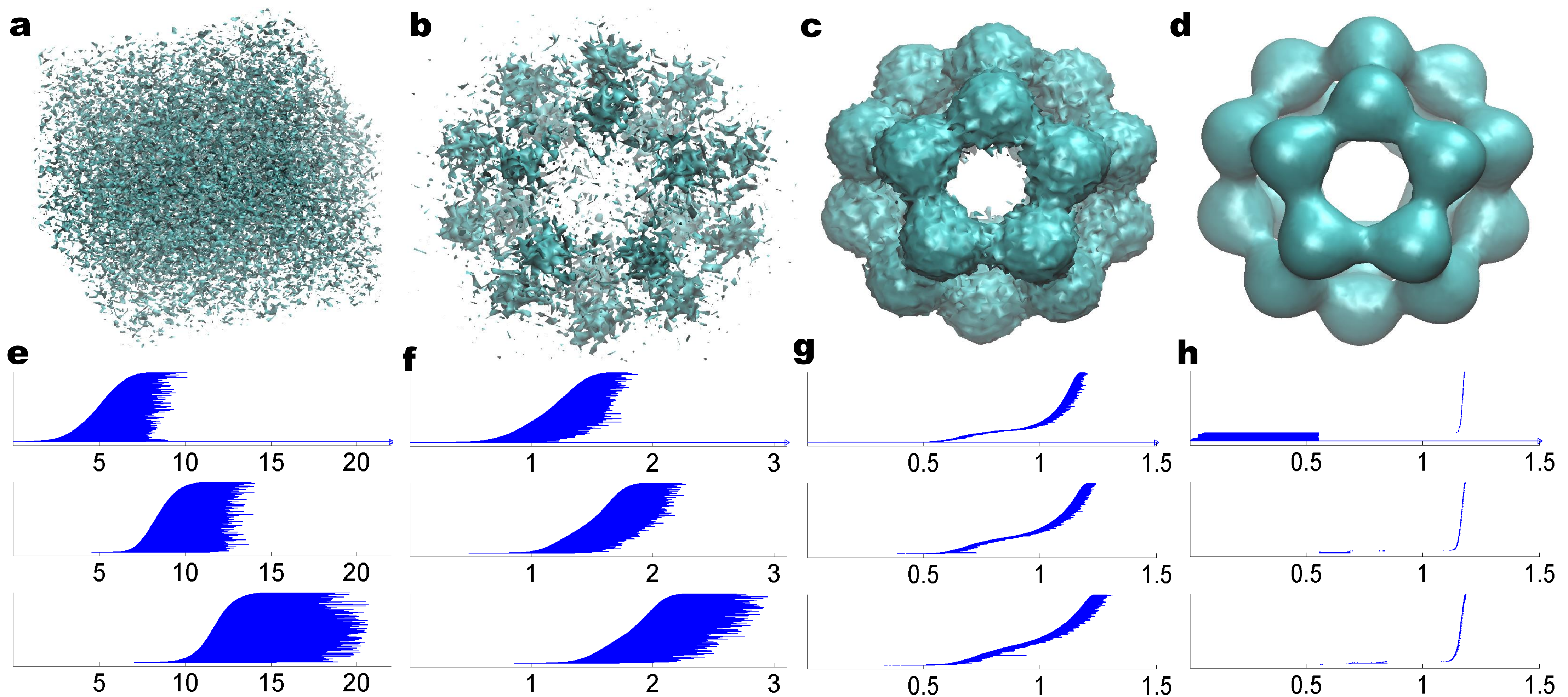}
\end{tabular}
\end{center}
\caption{The barcode representation for  contaminated fullerene C$_{20}$ data with different SNRs. The SNRs for {\bf a}, {\bf b}, {\bf c} and {\bf d} are 0.1, 1.0, 10.0 and 100.0, respectively,  and isovalues used for their visualization are 4.00, 1.00, 0.60, and 0.60, respectively. The corresponding barcodes are given in {\bf e}, {\bf f}, {\bf g} and {\bf h}, respectively.   The top, middle, and bottom panels are for the barcodes of $\beta_0$,  $\beta_1$ and $\beta_2$, respectively. It can be seen from the barcodes that when SNR is low, i.e., SNR $=0.1$ or 1.0, fullerene molecule is invisible. At a low noise level, molecular pattern emerges. More importantly, the persistence of the psudo-topological structure is directly related to the SNR. In the barcode representation, noise tends to induce a continuous band (or stripe) of bars, of which the width or relative persistent length is determined by the magnitude of the noise.   }
\label{fig:noise_c20}
\end{figure}

The density function described in Eq. (\ref{eq:rigidity3}) with the generalized exponential kernel shown in Eq. (\ref{eq:couple_matrix1}) is used to simulate the  density of fullerene C$_{20}$. We choose $\kappa=1.0$ and $\sigma=0.5$\AA~ in our study. Figure   \ref{fig:noise_c20} demonstrates the barcode representation for contaminated fullerene C$_{20}$ data with different SNRs. The SNRs for  Figs. \ref{fig:noise_c20} {\bf a}, {\bf b}, {\bf c} and {\bf d} are 0.1, 1.0, 10.0 and 100.0, respectively. It can be seen from  Figs. \ref{fig:noise_c20} {\bf a} and  {\bf b}, that when SNR is low, i.e., SNR$ =0.1$ or 1.0, fullerene atoms are invisible. When the SNR is increased in Figs. \ref{fig:noise_c20} {\bf c} and {\bf d}, the molecular intrinsic patterns begin to emerge.

\begin{figure}
\begin{center}
\begin{tabular}{c}
\includegraphics[width=0.9\textwidth]{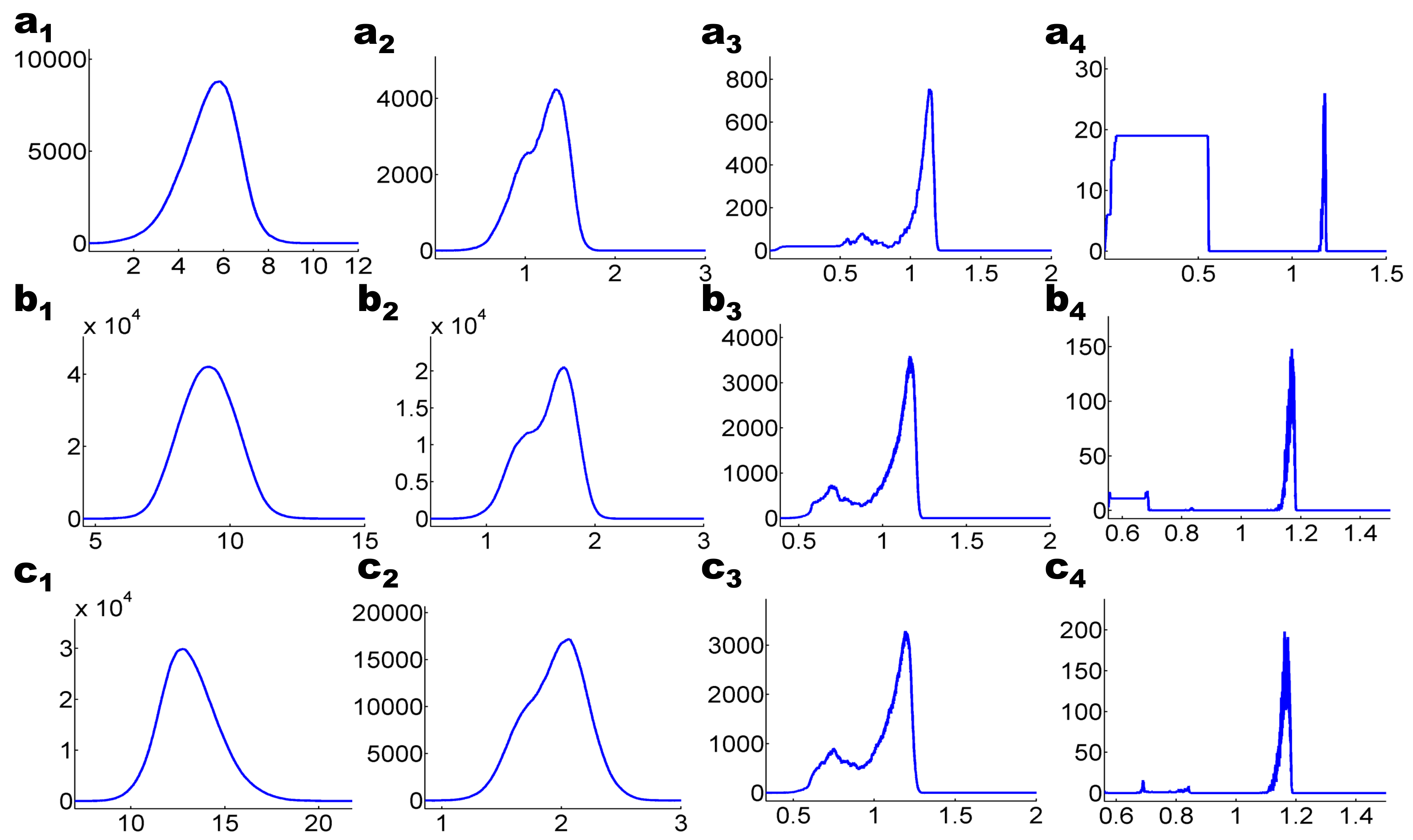}
\end{tabular}
\end{center}
\caption{The  histograms of topological invariants over the filtration process for  contaminated fullerene C$_{20}$ data with different SNRs. The  ${\bf a}_i$, ${\bf b}_i$ and ${\bf c}_i$ rows are the counts of  $\beta_0$, $\beta_1$ and $\beta_2$, respectively, where subscripts $i=1,2,3 $ and $4$ correspond to SNRs 0.1, 1.0, 10.0 and 100.0, respectively. It can be seen that as the SNR increases, barcodes for noise and signal gradually separate from each other. Since the Gaussian noise is used, the noise parts typically assume  Gaussian distributions in shape. }
\label{fig:noise_c20b}
\end{figure}

The topological signature of the above four cases can be analyzed  as shown in Figs.   \ref{fig:noise_c20} {\bf e}, {\bf f}, {\bf g} and {\bf h}. First, we note that all of three  topological invariants, namely $\beta_0$, $\beta_1$ and $\beta_2$, are very sensitive to and essentially dominated by noise. Additionally, noise gives rise to continuous bands (or stripes) of bars in all the three topological invariants. Moreover,  noise provides similarly numbers of bars in  $\beta_0$, $\beta_1$ and $\beta_2$ panels. Moreover, contradicting to the common belief that noise only contributes to short lived bars, the noise bars can be very long. The average band length of noise bars proportions  to  noise magnitude. In fact, in our recent work  \cite{KLXia:2014c}, we regard all bars, including short lived bars, of a protein as molecular topological fingerprints and as being of equal importance.  Furthermore, at low SNRs (i. e.,  Figs. \ref{fig:noise_c20} {\bf e} and {\bf f}),  the bars of three invariants maintain  Gaussian-like distributions with respect to the isovalue filtration (i. e., the $x$-axis) as shown in 
 Fig.   \ref{fig:noise_c20b}. However, at relatively high SNRs,  bars do not have the Gaussian-like distributions due to a relatively high C$_{20}$ density, see  Fig.   \ref{fig:noise_c20b}.

\begin{figure}
\begin{center}
\begin{tabular}{c}
\includegraphics[width=0.9\textwidth]{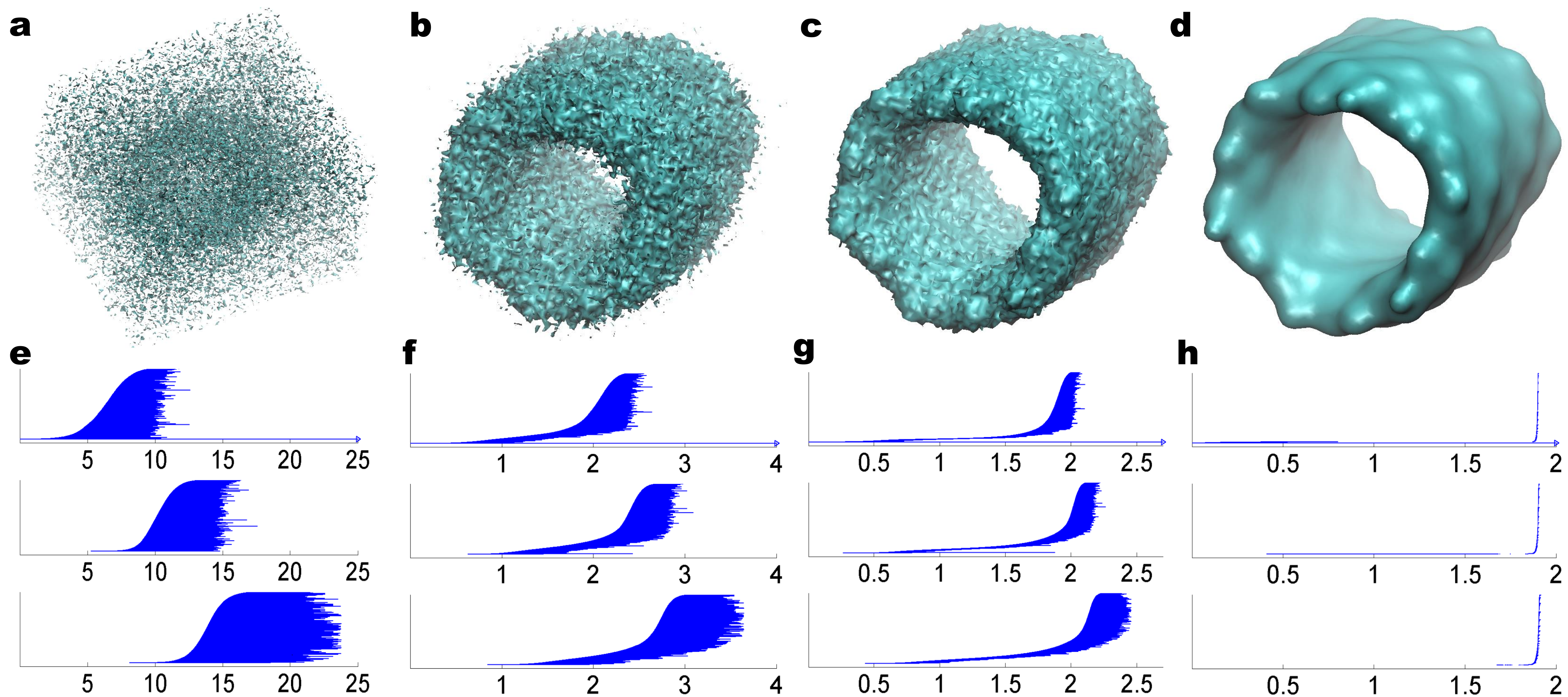}
\end{tabular}
\end{center}
\caption{The barcodes representation for protein 2GR8 beta segment with different SNRs. The SNRs for {\bf a}, {\bf b}, {\bf c} and {\bf d} are 0.1, 0.5, 1.0 and 10.0, respectively, and  isovalues used for their visualization are 5.00, 1.00, 1.00, and 1.00, respectively. The corresponding barcodes are presented in {\bf e}, {\bf f}, {\bf g} and {\bf h}, respectively.  The top, middle, and bottom panels are for the barcodes of $\beta_0$,  $\beta_1$ and $\beta_2$, respectively. It can be seen from the barcodes that when the SNR is large, i.e., SNR$=$0.1 or 0.5, the original topological properties are blurred. When the noise effect dwindles, the intrinsic patterns begin to emerge. More importantly, the persistence of the psudo-topological structure is directly related to the SNR. In the barcode representation, noise tends to induce a continuous band (or stripe) of bars, of which the width or relative persistent length is determined by the magnitude of the noise.  }
\label{fig:noise_2gr8}
\end{figure}

To further analyze  the topological signature  of noise, we consider a protein structure made of only beta sheets obtained from 2GR8. Again its total  density distribution is approximated by using the density function given in Eq. (\ref{eq:rigidity3}) realized by using the generalized exponential kernel of Eq. (\ref{eq:couple_matrix1}) with $\kappa=1.0$ and $\sigma=2.0$\AA. Figure  \ref{fig:noise_2gr8} shows our results. The SNRs for Figs. \ref{fig:noise_2gr8}  {\bf a}, {\bf b}, {\bf c} and {\bf d} are 0.1, 0.5, 1.0 and 10.0, respectively. The corresponding barcodes are given in Figs. \ref{fig:noise_2gr8}  {\bf e}, {\bf f}, {\bf g} and {\bf h}.  The  topological signature of noise in Fig. \ref{fig:noise_2gr8} is quite similar to that in  Fig.  \ref{fig:noise_c20b}, Essentially,   Gaussian noise induces  continuous bands (or stripes) of bars, whose  width or relative persistent length is determined by noise intensity.

As demonstrated in above examples, barcodes present a topological description of Gaussian noise. The related band width of the noise can be used to assess noise magnitude. This qualitative description can be further used as a guidance for topological noise reduction and topological fingerprint identification.

\subsection{Topological denoising}\label{Sec:HighOrderGF2}

\paragraph{Geometric flows}
Geometric PDEs offer an efficient approach for noise reduction. High order geometric PDEs were first introduced for edge-preserving image restoration in 1999 and have a general form \cite{Wei:1999}
\begin{eqnarray}\label{eqn:highorder}
\frac{\partial u ({\bf r},t)}{\partial t}  =- \sum_{q}\nabla \cdot {\bf j}_q
+ e (u ({\bf r},t),|\nabla u ({\bf r},t)|, t), \quad q=0,1,2,\cdots
\end{eqnarray}
where the nonlinear hyperflux term ${\bf j}_q$ is given by
\begin{eqnarray}\label{eqn:highorder1-1}
{\bf j}_q=-
d_q (u ({\bf r},t),|\nabla u ({\bf r},t)|, t) \nabla \nabla^{2q}
u ({\bf r},t), \quad
q=0,1,2,\cdots
\end{eqnarray}
where ${\bf r}\in {\mathbb R}^n$, $ \nabla=\frac{\partial}{\partial{\bf r}} $, $u ({\bf r},t)$ is the processed image function,
$d_q (u ({\bf r},t),|\nabla u ({\bf r},t)|, t)$ are edge sensitive diffusion coefficients
and $e (u ({\bf r},t),|\nabla u ({\bf r},t)|, t)$  is a nonlinear operator.
The original noise data $X ({\bf r})$ is used as the initial input $u ({\bf r},0)=X ({\bf r})$.  The hyper-diffusion coefficients  $d_q (u,|\nabla u|,t)$ in Eq.  (\ref{eqn:highorder1-1}) can also be chosen as the Gaussian form
\begin{eqnarray}\label{eqn:highorder2}
d_q (u ({\bf r},t),|\nabla u ({\bf r},t)|,
t)=d_{q0}\exp\left[-\frac{|\nabla u|^2}{2\sigma_q^2} \right],
\end{eqnarray}
where $d_{q0}$ is chosen as a constant with value depended on the noise level, and $\sigma_0$ and $\sigma_1$ are local statistical variance of $u$ and $\nabla u$
\begin{eqnarray}\label{eqn:highorder3}
\sigma_q^2 ({\bf r})= \overline{|\nabla^q u -\overline{\nabla^q
u}|^2} \quad  (q=0,1).
\end{eqnarray}
Here the notation $\overline{Y ({\bf r})}$ represents the local average of $Y ({\bf r})$ centered at position ${\bf r}$.

High order geometric PDEs have many practical applications  \cite{Wei:1999,MLysaker:2003,GGilboa:2004}. They have been specifically modified for molecular surface formation and evolution  \cite{Bates:2009} as,
 \begin{equation}\label{model4th3}
\frac{\partial S}{\partial t} = (-1)^q \sqrt{g ( |\nabla\nabla^{2q}
S|)} \nabla \cdot \left ( \frac{ \nabla  (\nabla^{2q} S)} { \sqrt{g (
|\nabla\nabla^{2q} S |)}} \right) + P (S,|\nabla S|),
\end{equation}
where $S$ is the hypersurface function,  $g (|\nabla\nabla^{2q}S|)=1+ |\nabla\nabla^{2q} S|^2$ is the generalized Gram determinant and $P$ is a generalized potential term. When $q=0$ and $P=0$, a Laplace-Beltrami equation is obtained \cite{Bates:2008},
 \begin{equation}\label{mean curvature_flow}
\frac{\partial S}{\partial t} =   |\nabla S| \nabla \cdot \left ( \frac{ \nabla S} {
|\nabla S |} \right).
\end{equation}
We employ this Laplace-Beltrami equation for  the noise reduction in this paper.

\paragraph{Topological fingerprint identification}\label{Sec:Topological_fingerprint}
\begin{figure}
\begin{center}
\begin{tabular}{c}
\includegraphics[width=0.9\textwidth]{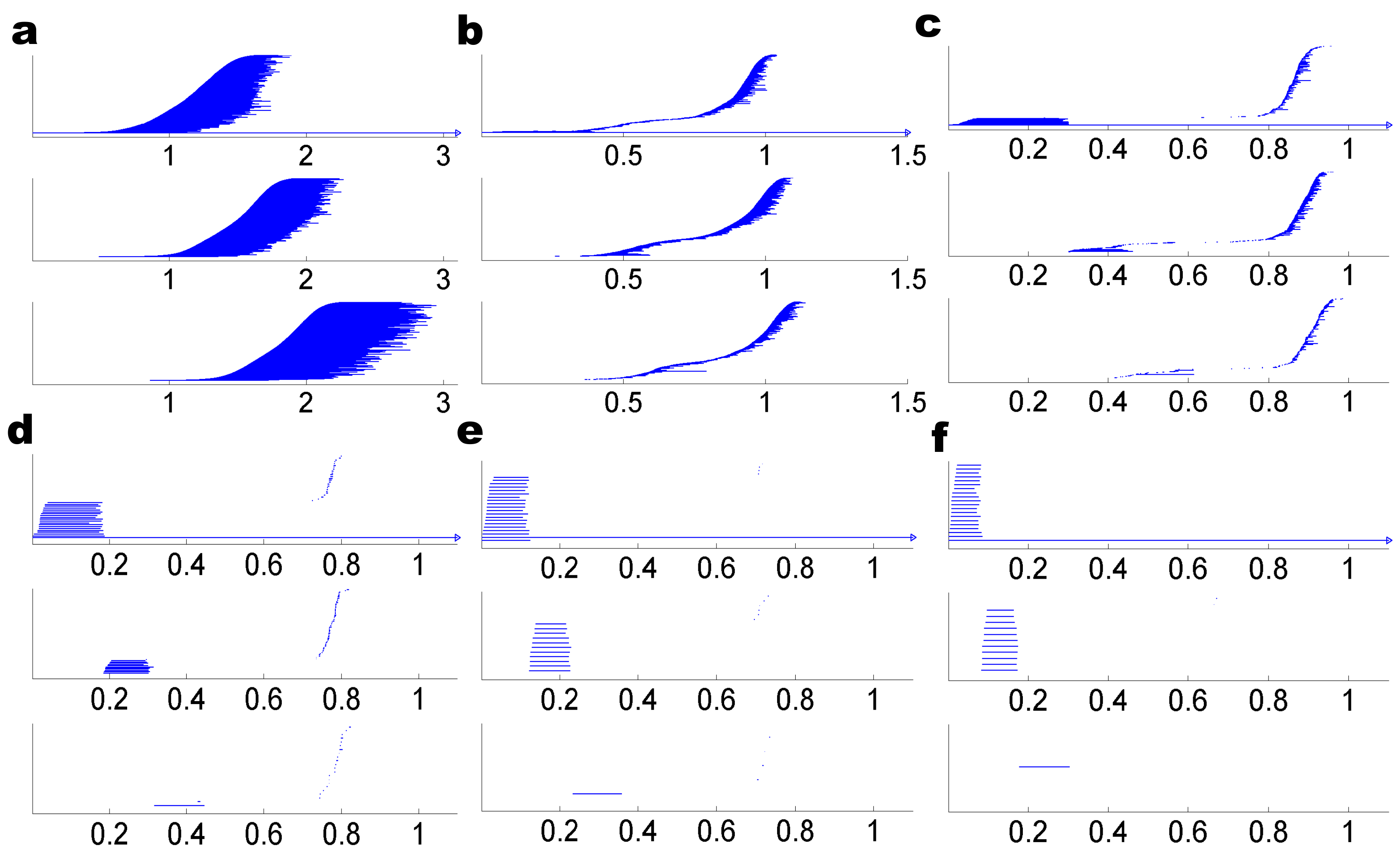}
\end{tabular}
\end{center}
\caption{The barcodes representation for denoising contaminated fullerene C$_{20}$ with SNR 1.0. The barcodes for fullerene C$_{20}$ with SNR 1.0 is demonstrated in {\bf a}. The denoising steps for {\bf b}, {\bf c}, {\bf d}, {\bf e} and {\bf f} are 20, 40, 60, 80 and 100, respectively. The noise induced topological invariants have been gradually weakened and finally eradicated. Compared with the original noise-polluted barcodes in {\bf a}, the noise effect has been enormously scaled down after only 20 steps of denoising as indicated in {\bf b}. In {\bf c}, there is a clear separation between the intrinsic topological features of fullerene C$_{20}$ and noise induced topological invariants.  From {\bf d} to {\bf f}, the noise effect is further reduced, and we are able to identify a persistent barcode pattern, which is an indication of the intrinsic topological invariants of fullerene C$_{20}$. It should also be noticed that the denoising process fades noise intensity, therefore noise induced topological patterns gradually move to the right of filtration parameter and finally disappear. }
ants have been gradually weakened and finally eradicated. Compared with the original noise-polluted barcodes in {\bf a}, the noise effect has been enormously scaled down after only 20 steps of denoising as indicated in {\bf b}. In {\bf c}, there is a clear separation between the intrinsic topological features of fullerene C$_{20}$ and noise induced topological invariants.  From {\bf d} to {\bf f}, the noise effect is further reduced, and we are able to identify a persistent barcode pattern, which is an indication of the intrinsic topological invariants of fullerene C$_{20}$. It should also be noticed that the denoising process fades noise intensity, therefore noise induced topological patterns gradually move to the right of filtration parameter and finally disappear.
\label{fig:denoise_c20}
\end{figure}

Computationally, the finite different method is used to discretize Eq. (\ref{mean curvature_flow}). Suitable time interval $\Delta t$, and grid spacing $h$ are needed. For cryo-EM data, its voxel spacing is related to the data resolution and varies greatly. For example, The voxel spacings of EMD1776, EMD1229 and EMD5729  are 1.69 \AA, ~ 4.00 \AA,~ and 4.16 \AA, respectively. In our simulated fullerene C$_{20}$ , C$_{60}$ and protein 2GR8 examples, the voxel spacings are 0.06 \AA, 0.08 \AA, and 0.40 \AA, respectively. To avoid confusion and control the noise reduction process systematically, we simply ignore the voxel spacing and use the unified parameters $\Delta t=1.0E-5$ and $h=1.0E-2$. The intensity of denoising is then described by the number of iteration steps.

\begin{figure}
\begin{center}
\begin{tabular}{c}
\includegraphics[width=0.9\textwidth]{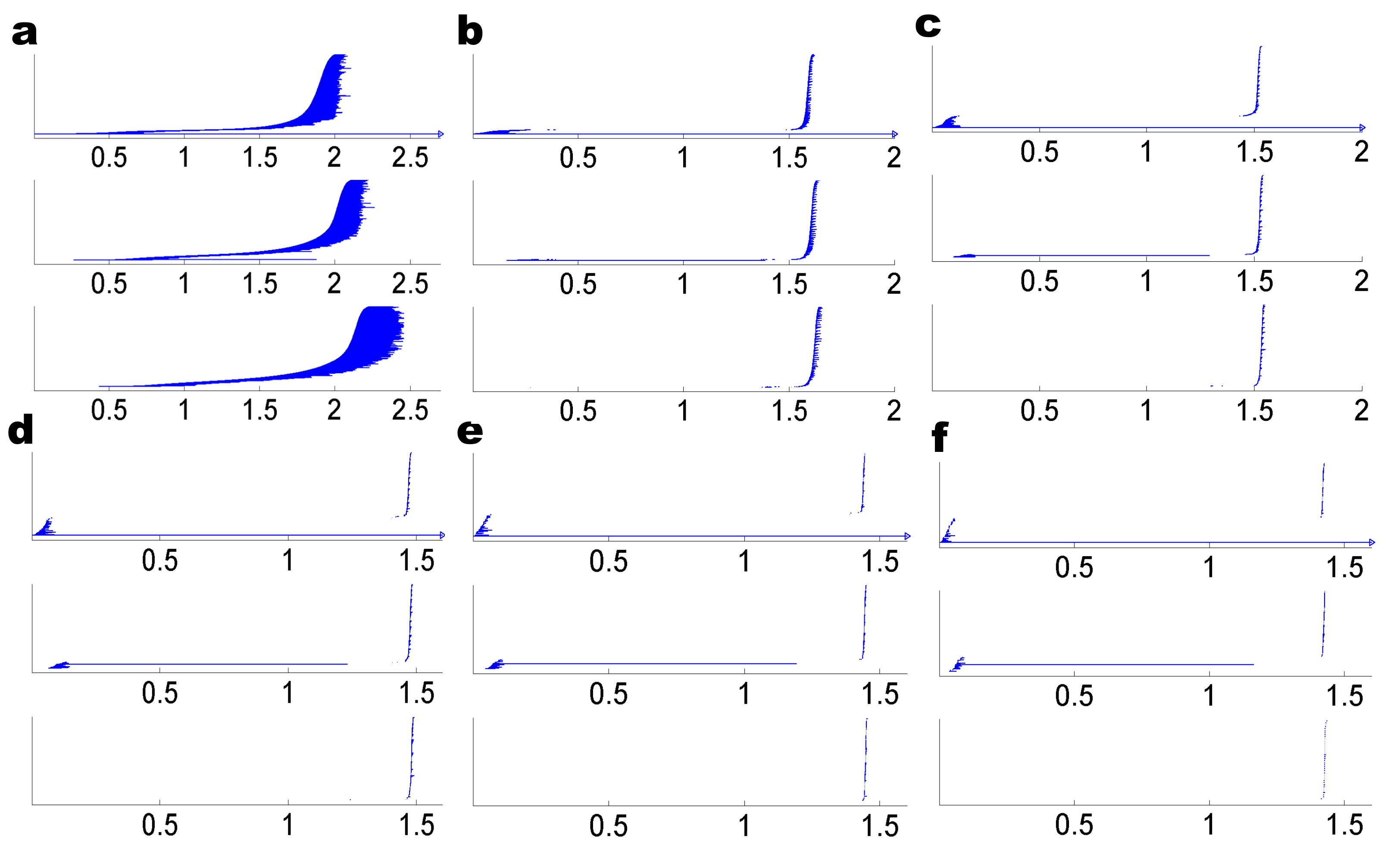}
\end{tabular}
\end{center}
\caption{The barcodes representation for the noise reduction of contaminated protein 2GR8 beta segment with SNR 1.0. The barcodes for 2GR8 with SNR 1.0 is demonstrated in {\bf a}. The denoising steps for {\bf b}, {\bf c}, {\bf d}, {\bf e} and {\bf f} are 10, 20, 30, 40 and 50, respectively. In this case, the noise induced topological invariants have been gradually weakened but not eradicated. Compared with the original noise-polluted barcodes in {\bf a}, the noise effect has been enormously scaled down after 10 steps of denoising as indicated in {\bf b}. From {\bf c} to {\bf f}, there is a clear separation between the intrinsic topological features of protein segment and noise induced topological invariants. The noise effect is continuously reduced, and we are able to identify a persistent barcode pattern, which is an indication of the intrinsic topological invariants of the protein segment. However, unlike the fullerene C$_{20}$, further denoising of 2GR8 data  will remove both the noise  and intrinsic structure related topological information.  }
\label{fig:denoise_2gr8}
\end{figure}

The noise reduction effectiveness is commonly validated by a visual comparison with the original results. Quantitative assessment usually proves to be difficult, as noise and signal or image information can be tightly entangled. In this section, we propose a topological method for monitoring the evolution of relative behaviors of noise and signal during the denoising process. More specifically, persistent barcodes from a series of denoising data are compared. The noise signature and topological  fingerprint  in these barcodes are carefully studied. The present emphasis is on the topological fingerprint identification.

\begin{figure}
\begin{center}
\begin{tabular}{c}
\includegraphics[width=0.9\textwidth]{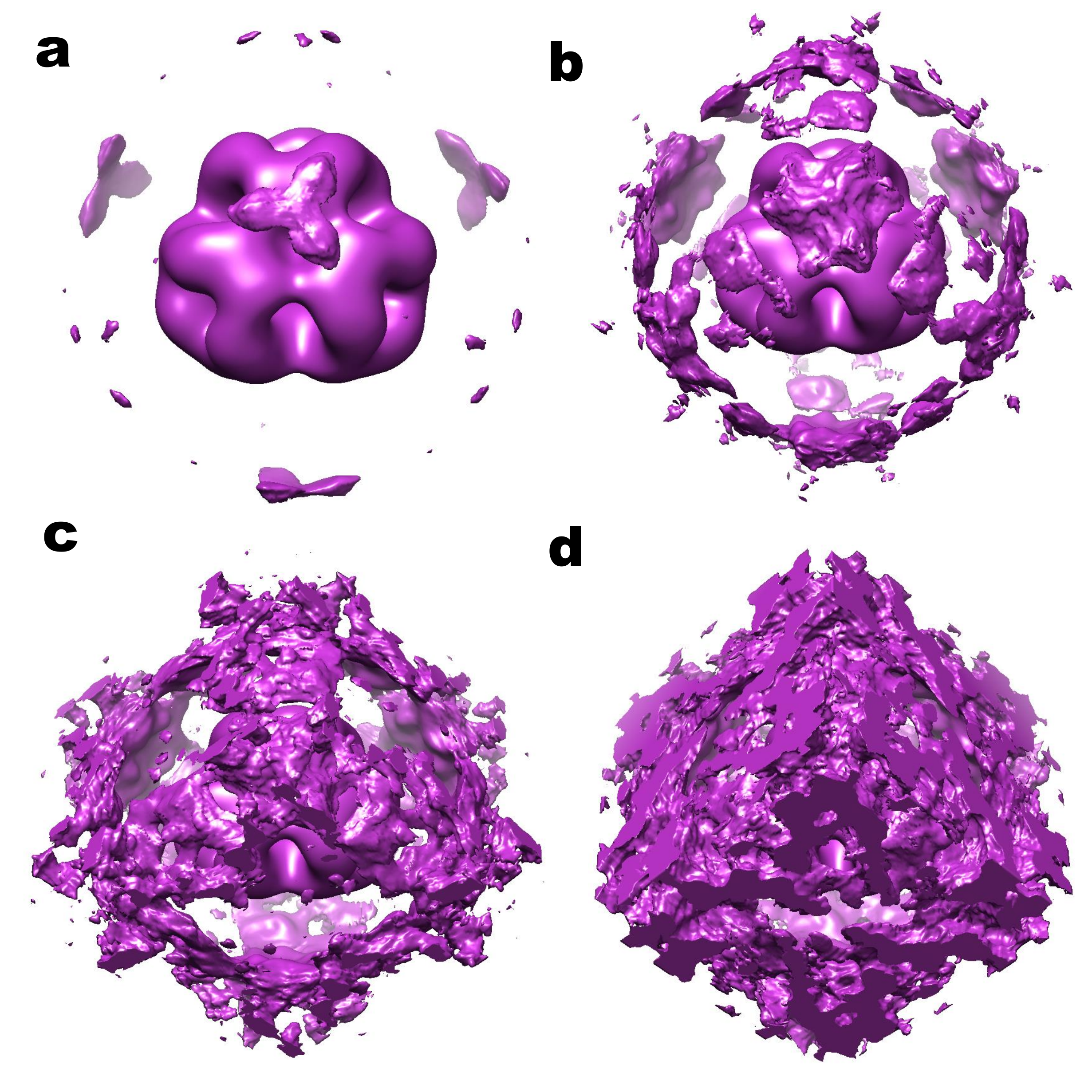}
\end{tabular}
\end{center}
\caption{The original noise in EMD 1776 data. The isovalues for {\bf a}, {\bf b}, {\bf c} and {\bf d} are 0.020, 0.010, 0.005, and 0.000, respectively. }
\label{fig:noise_1776}
\end{figure}

During the denoising process, topological features of both signal and noise are constantly evolving. To quantitatively analyze their behaviors, three cases with Gaussian white noise  and a case with Cryo-EM data are considered. The first case is a noise contaminated fullerene C$_{20}$ with SNR 1.0. The persistent barcode results are demonstrated in Fig.  \ref{fig:denoise_c20}.  We vary the number of denoising steps in our study. For Figs.  \ref{fig:denoise_c20}{\bf b}, {\bf c}, {\bf d}, {\bf e} and {\bf f}, the numbers of iterations are 20, 40, 60, 80 and 100, respectively. The noise induced topological persistence has been gradually weakened and finally cleaned up. Compared with the original noise-polluted barcodes in Fig.  \ref{fig:denoise_c20}{\bf a}, the noise effect has been enormously scaled down after only 10 steps of denoising as indicated in Fig.  \ref{fig:denoise_c20}{\bf b}. In Fig.  \ref{fig:denoise_c20}{\bf c}, there is a clear separation between the intrinsic topological persistence of fullerene C$_{20}$ and noise induced topological persistence.  From Fig.  \ref{fig:denoise_c20}{\bf d} to Fig.  \ref{fig:denoise_c20}{\bf f}, the noise effect is further weakened, and thus we are able to identify a consistent barcode pattern, which is an indication of the intrinsic topological invariants of fullerene C$_{20}$. It should also be noticed that since noise intensity diminishes  during the denoising process,  noise induced topological patterns gradually shift to the right of the filtration parameter and eventually disappear.

The second case is a noise contaminated protein segment from 2GR8 with SNR 1.0. Its topological behavior under  the denoising process is illustrated in Fig.  \ref{fig:denoise_2gr8}. In this case, the noise induced topological invariants have been gradually weakened but not eradicated. Compared with the original noise-polluted barcodes in Fig.  \ref{fig:denoise_2gr8}{\bf a}, the noise persistence has been enormously reduced after 10 steps of denoising as indicated in Fig.  \ref{fig:denoise_2gr8}{\bf b}. From Fig.  \ref{fig:denoise_2gr8}{\bf b} to Fig.  \ref{fig:denoise_2gr8}{\bf f}, there is a clear separation between the intrinsic topological invariants of protein segment and noise induced topological invariants. As the noise effect is continuously weakened, we are able to identify a persistent barcode pattern, which is an indication of the intrinsic topological invariants of the protein segment. However, unlike the situation for fullerene C$_{20}$, further denoising will remove both the noise  and intrinsic structure related topological information.

\begin{figure}
\begin{center}
\begin{tabular}{c}
\includegraphics[width=0.9\textwidth]{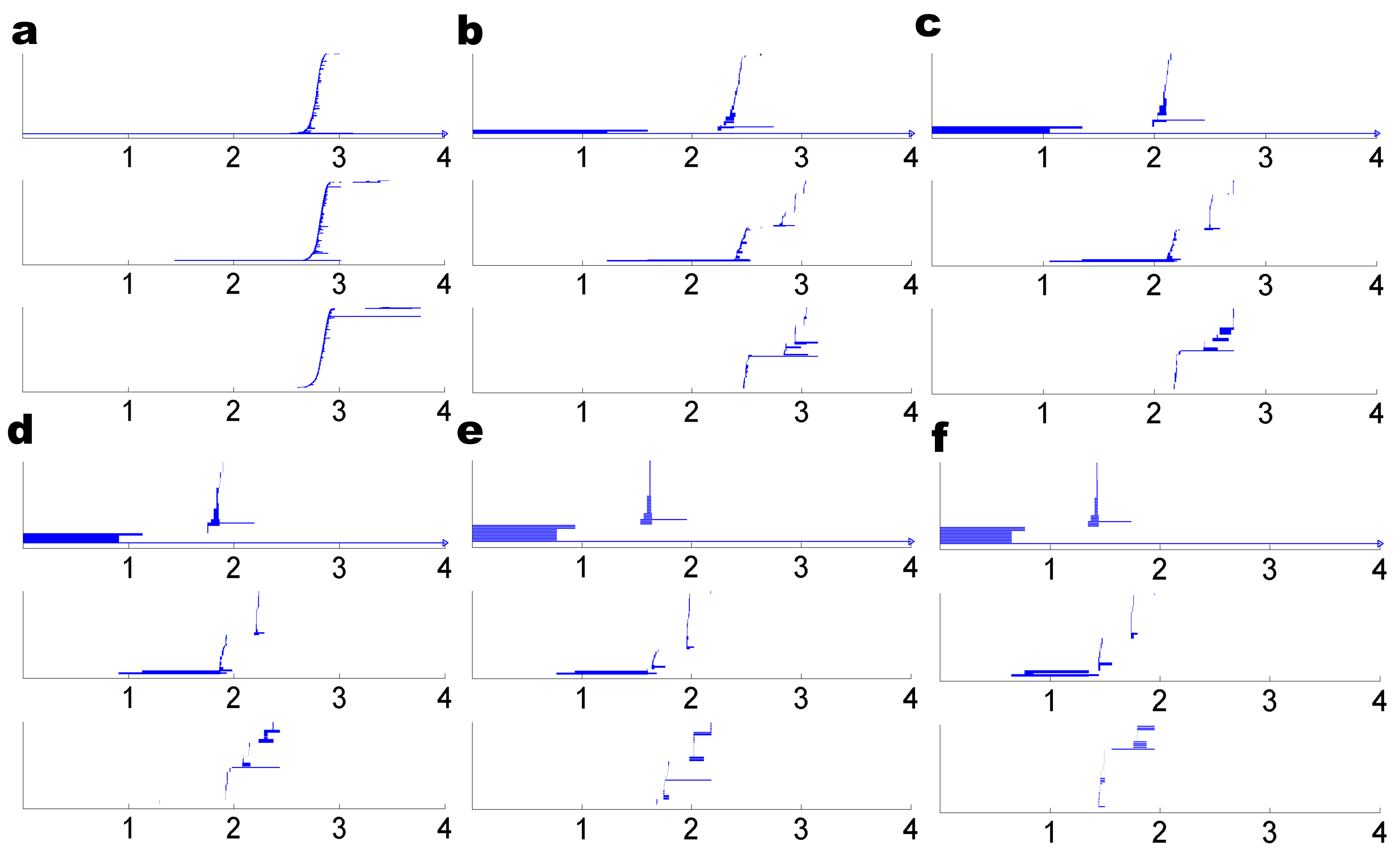}
\end{tabular}
\end{center}
\caption{The barcodes representation for the noise removal of contaminated EMD 1776 with SNR 1.0. The denoising steps for {\bf a}, {\bf b}, {\bf c}, {\bf d}, {\bf e} and {\bf f} are 10, 40, 80, 120, 160 and 200, respectively. In this case, the noise induced topological invariants have been gradually weakened but not eradicated. Compared with the original noise-polluted barcodes in Fig. \ref{fig:noise_2gr8} {\bf b}, the noise effect has been enormously reduced after 10 steps of denoising as indicated in {\bf a}. From {\bf b} to {\bf f}, there is a clear separation between the intrinsic topological features of protein segment and noise induced topological invariants. The noise effect  continuously wanes, and we are able to identify a persistent barcode pattern, which is an indication of the intrinsic topological invariants of the protein segment. However, unlike the fullerene C$_{20}$, further denoising will remove both the noise  and intrinsic structure related topological information.}
\label{fig:denoise_1776}
\end{figure}

Finally, we consider a more realistic example, i.e., EMD 1776, obtained from cryo-EM. For EMD 1776, when the isovalue goes to around 0.020, noise begins to emerge. Figure  \ref{fig:noise_1776} depicts noise in EMD 1776 data. The isovalues for Figs.  \ref{fig:noise_1776} {\bf a}, {\bf b}, {\bf c},  and {\bf d} are 0.020, 0.010, 0.005 and 0.000, respectively. The geometric flow based denoising method is employed.  Persistent homology results are demonstrated in Fig.  \ref{fig:denoise_1776}.  The numbers of denoising steps in Figs. \ref{fig:denoise_1776} {\bf a}, {\bf b}, {\bf c}, {\bf d}, {\bf e} and {\bf f} are 10, 40, 80, 120, 160 and 200, respectively. In this case, the noise induced topological invariants have been gradually weakened but have not been cleaned up.

From the above analysis, some common features can be unveiled. First, signal related topological features tend to be buried near the left end of the filtration parameter during the denoising process. Second, topological features corresponding to signal and noise begin to separate as the denoising procedure advances. Third,   intrinsic topological invariants associated with the ``signal'' of the data are essentially  preserved over the denoising process.  These features can be used to guide the evaluation and thresholding of the denoising process.

\begin{figure}
\begin{center}
\begin{tabular}{c}
\includegraphics[width=0.9\textwidth]{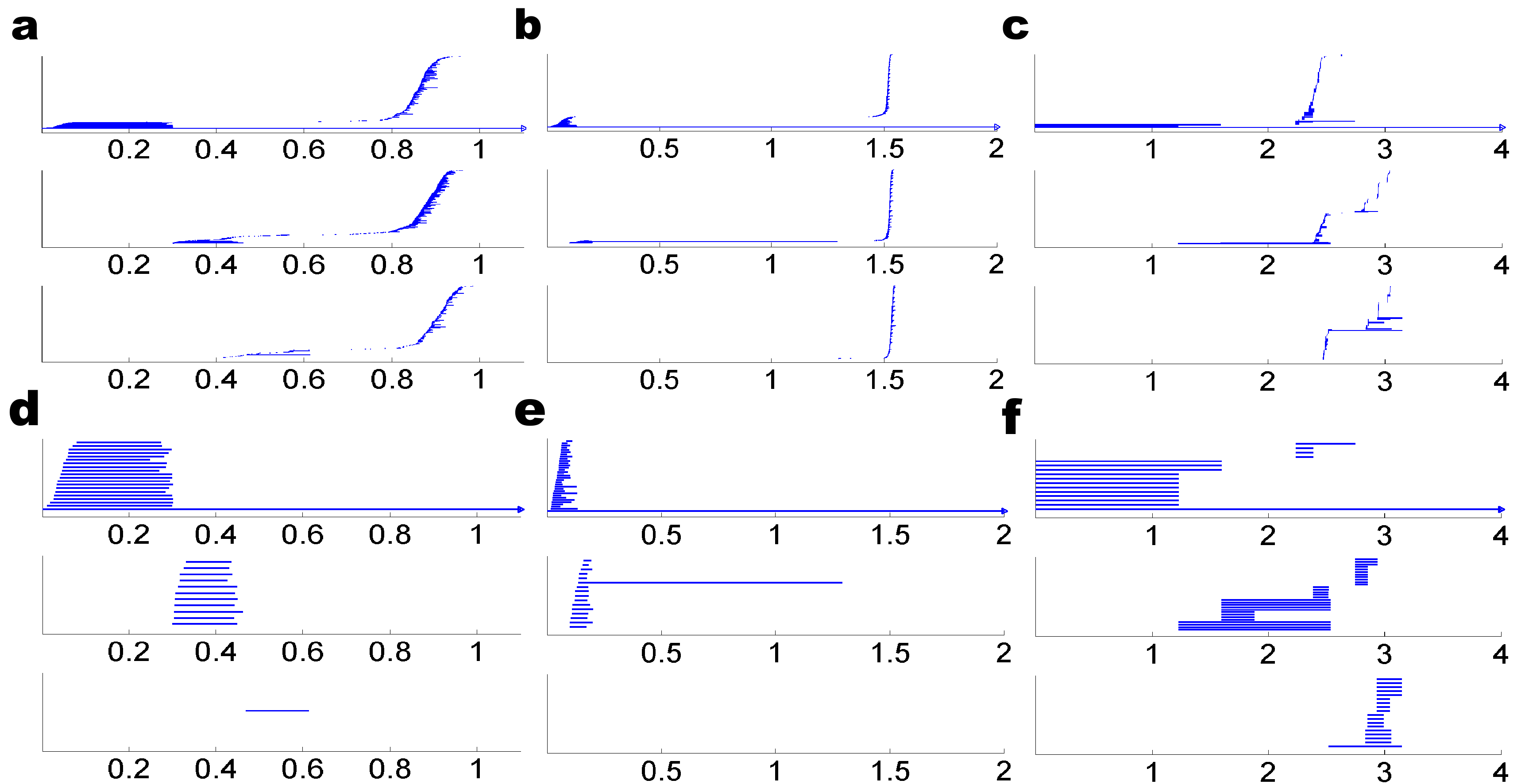}
\end{tabular}
\end{center}
\caption{Retrieve barcode patterns through noise thresholding for three systems. Subfigure  {\bf a} is the barcode for denoising fullerene C$_{20}$ data with 20 iterations as in Fig. \ref{fig:denoise_c20} {\bf b}. Subfigure  {\bf b} is the barcode for denoising protein 2GR8 beta segment data with 20 iterations as in Fig. \ref{fig:denoise_2gr8} {\bf c}.  Subfigure {\bf c} is the barcode for denoising EMD 1776 data with 40 iterations as in Fig.  \ref{fig:denoise_1776} {\bf b}. Barcodes in {\bf d}, {\bf e} and {\bf f} are all obtained respectively  from {\bf a}, {\bf b} and {\bf c}  by setting noise threshold as 0.1, i.e., removing all barcodes with their lengths shorter than 0.1. }
\label{fig:noise_threshold}
\end{figure}

When the persistent features in the barcode representation is identified, one can retrieve the intrinsic barcodes due to the signal by simply setting up a noise threshold and removing all barcodes with length less than it \cite{Fasy:2013}. Figure  \ref{fig:noise_threshold} demonstrates this technique. Figure \ref{fig:noise_threshold}{\bf a} shows the  barcodes after 20 iterations  of the noisy fullerene C$_{20}$ data as given in Fig. \ref{fig:denoise_c20} ({\bf b}). Figure  \ref{fig:noise_threshold} {\bf b} illustrates the barcodes after 20 iterations of the noisy  protein 2GR8 beta segment data  as given in Fig. \ref{fig:denoise_2gr8} ({\bf c}). Figure \ref{fig:noise_threshold} {\bf c} depicts the  barcodes after 40 iterations of  the noisy EMD1776 data as in presented Fig.  \ref{fig:denoise_1776} {\bf b}. Figures \ref{fig:noise_threshold} {\bf d}, {\bf e} and {\bf f} display the barcodes of the above three cases with a noise threshold of 0.1, i.e., removing all barcodes with their lengths shorter than 0.1.

Another importance aspect is that for cryo-EM data, we just need to consider voxels with isovalue larger than a certain threshold. For instance, in EMD 1776 data,  noise begins to emerge when the isovalue goes down to about 0.020 as indicated in Fig. \ref{fig:noise_1776}. As the isovalue decreases, more noise emerges. More importantly, in most cryo-EM data, isovalue can even go below 0.0. These special voxels, as far as we know, do not really represent the desirable biomolecular structure or more specifically do not directly reflect the desirable biomolecular structure. Usually, a recommended isovalue is specified for each cryo-EM data. The meaningful structure information can be only derived from voxels with value near this specified isovalue. From our persistent analysis, we believe that all the voxels with isovalue larger than certain threshold can be related to their inner structure properties. In order not to overlook certain potential structure pattern, in our persistent analysis, we just ignore all voxels with isovalue smaller than 0.0. This is usually done by assigning all negative isovalues  to 0.0.

\subsection{Case study:  EMD 5729 }

\begin{figure}
\begin{center}
\begin{tabular}{c}
\includegraphics[width=0.9\textwidth]{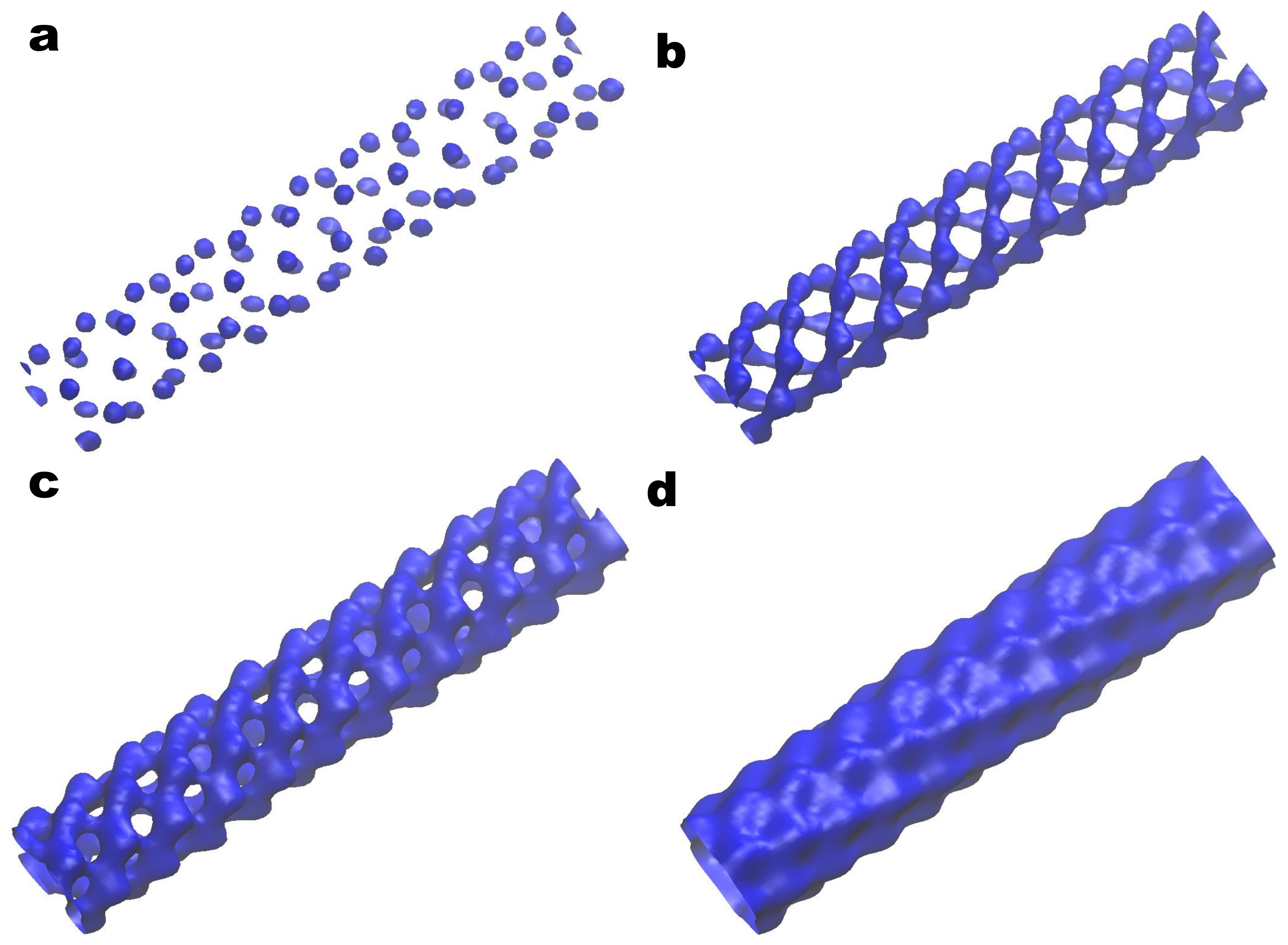}
\end{tabular}
\end{center}
\caption{Isosurfaces of EMD 5729 data after 20 steps of noise  reduction.  Images in {\bf a}, {\bf b}, {\bf c} , and {\bf d} are extracted from isovalues 0.35, 0.30, 0.25, and 0.10, respectively.  }
\label{fig:5729_surface}
\end{figure}

Finally, we consider  EMD 5729,  a supramolecular filamentous complex \cite{Qiao:2013},  to demonstrate the application of topological denoising  in cryo-EM data analysis. We first process the EMD 5729 data with 20 steps of noise reduction using our geometric flow method. The resulting data are illustrated in Fig.   \ref{fig:5729_surface} with four different isovalues.

\begin{figure}
\begin{center}
\begin{tabular}{c}
\includegraphics[width=0.9\textwidth]{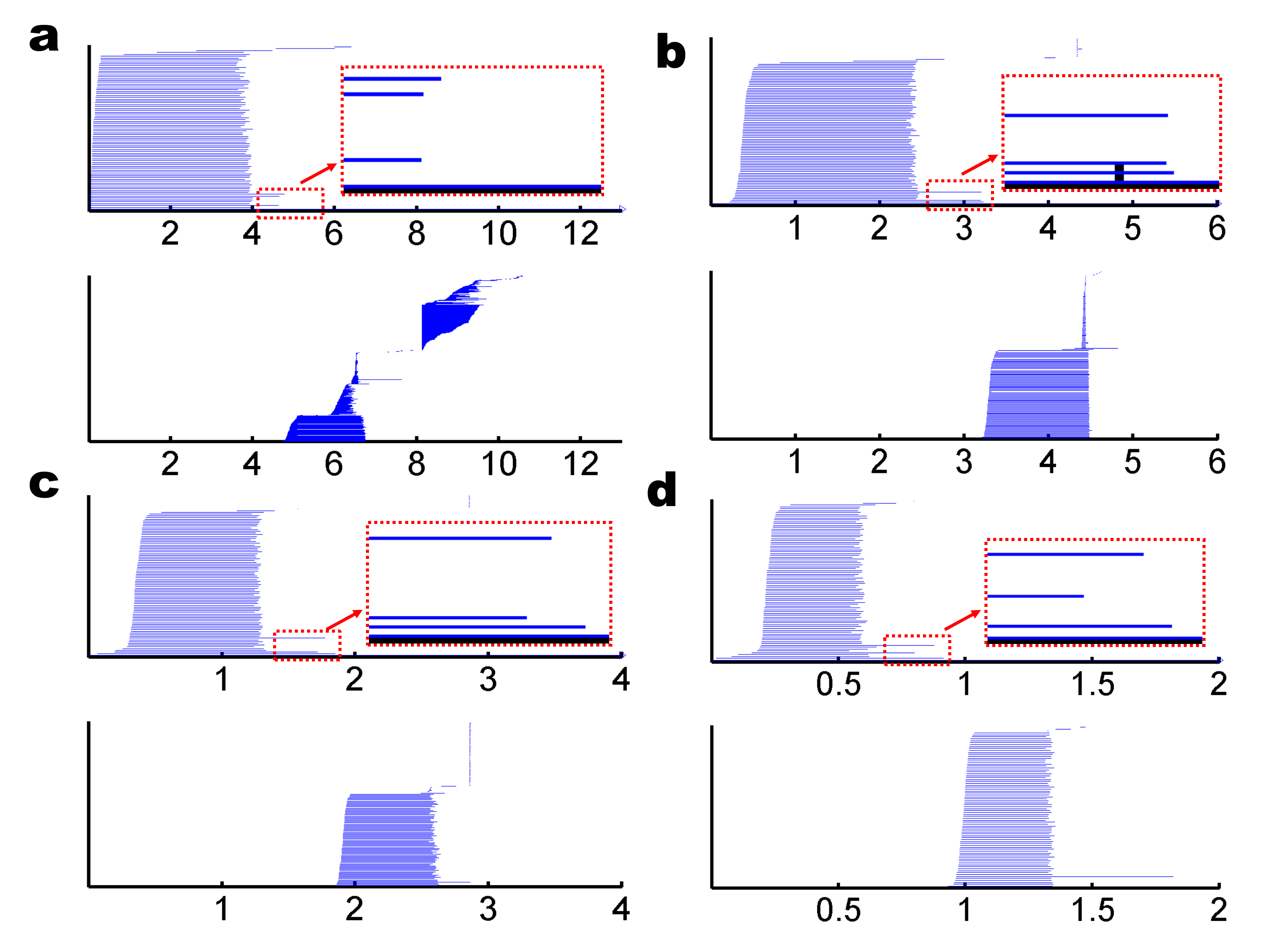}
\end{tabular}
\end{center}
\caption{The $\beta_0$ and $\beta_1$ barcodes for EMD 5729 in the noise reduction process. The barcodes for the original data are shown in {\bf a}. Here  {\bf b}, {\bf c} , and {\bf d} are barcodes after 10, 20 and 30 steps of noise reduction, respectively. Four relatively long $\beta_0$ bars in the highlighted circles indicate that there are four polymer strands in the structure. }
\label{fig:5729_barcodes}
\end{figure}

Figure \ref{fig:5729_barcodes} depicts barcodes computed for  EMD 5729. The $\beta_0$   pattern in  Fig. \ref{fig:5729_barcodes}{\bf a} shows a large number of bars of similar  lengths, which  indicates  only one type of protein monomers. Four relatively long persistent bars in the highlighted circle  indicate that there are four major pieces in the structure. The $\beta_1$ panel appears to be heavily contaminated by noise, which suggests the necessity for a denoising process.  The denoised $\beta_0$ topological persistent patterns  in  Figs. \ref{fig:5729_barcodes} {\bf b}, {\bf c} and {\bf d} confirm that only one type of bars can be found, which means there is only one type of polymer monomers. Additionally,  four relatively long $\beta_0$ bars confirm there are four polymer chains. The $\beta_1$ bars in  Figs. \ref{fig:5729_barcodes} {\bf b}, {\bf c} and {\bf d} are relatively consistent over the denoising process and have very similar lengths, which suggest these polymers are evenly distributed and form certain tunnel type of global structures with high symmetry. The long $\beta_1$ bar in Fig. \ref{fig:5729_barcodes}   {\bf d} indicates a large cylinder structure. Indeed, Fig.   \ref{fig:5729_surface} shows that protein monomers form four helix polymers and then bind together to result in a hollow cylinder structure \cite{Qiao:2013}.

\section{Persistent homology analysis of  microtubule   } \label{Sec:PH_microtubule}

In this section, persistent homology and topological denoising approaches are applied to the microtubule cryo-EM data analysis.

\subsection{Microtubule structure EMD-1129}

Microtubule is a cytoskeleton component of eukaryotic cells. It plays important roles in maintaining the structure or shape of the cell, supporting intracellular transport and  facilitating cell division  (mitosis and meiosis)  \cite{Nogales:2006}. Microtubule has a long hollow cylinder structure made up of polymerized $\alpha$- and $\beta$- tubulin dimers. These hetero-dimers bind head to tail into protofilaments, which further combine with each other in a parallel manner. The hollow cylinder structure of microtubule is finalized by attach about 13 protofilaments with each other side by side. Although the crystal structures of $\alpha-$ and $\beta-$ tubulin are available, experimental microtubule structure data are usually in low resolution and inadequate to separate  between neither $\alpha-$ tubulin and $\beta-$ tubulin, nor intra-dimer interface and inter-dimer interface. Recently, a microtubule intermediate structure data  (EMD-1129) with a 12 Angstrom resolution is obtained \cite{HWWang:2005}. Based on these data, we demonstrate how can we make use of our persistent homology and topological denoising methods to aid the modeling of cryo-EM  structures.

\begin{figure}
\begin{center}
\begin{tabular}{c}
\includegraphics[width=0.8\textwidth]{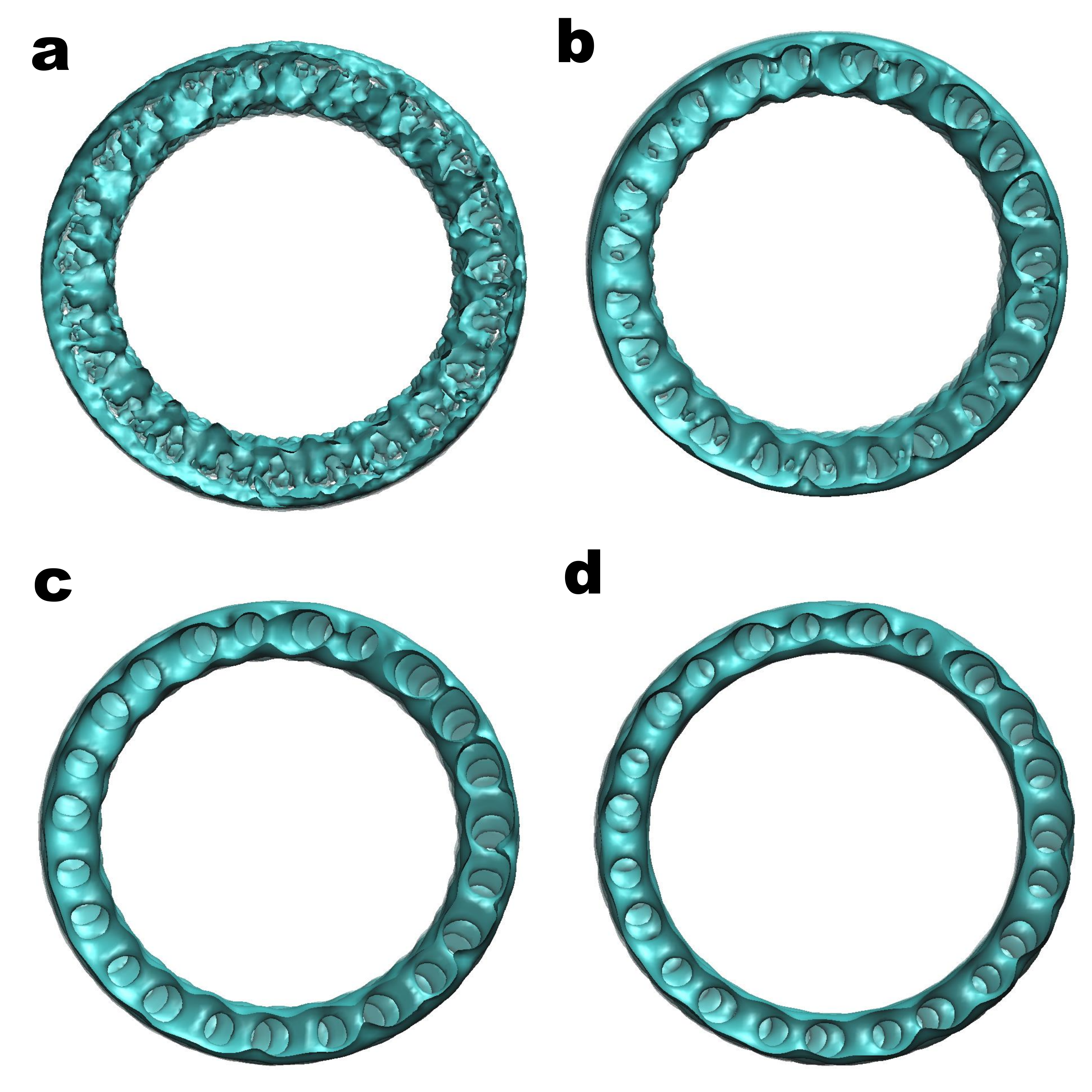}
\end{tabular}
\end{center}
\caption{EMD 1129 data preprocessing. {\bf a}: Surface extracted from original data with isovalue 16. {\bf b}, {\bf c} and {\bf d} are surfaces extracted from denoising data with 10, 20, and 40 iterations, respectively. }
\label{fig:EMD1129}
\end{figure}

\subsection{Coarse-grained models for microtubule  }\label{Sec:three_models}
\begin{figure}
\begin{center}
\begin{tabular}{c}
\includegraphics[width=0.8\textwidth]{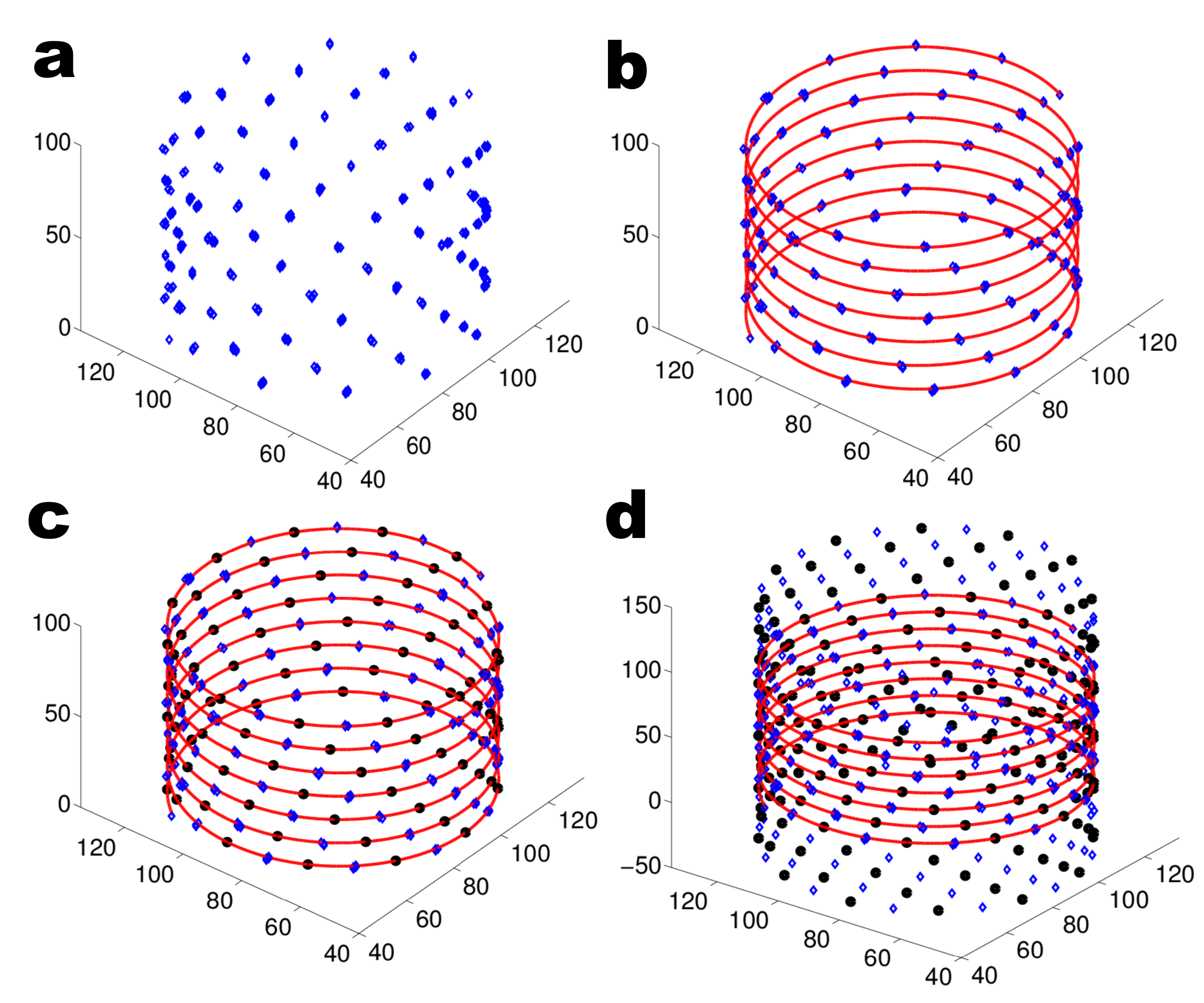}
\end{tabular}
\end{center}
\caption{The helix backbone reconstruction for the theoretical  model with two types of protein monomers. In {\bf a}, positions of voxels with isovalue larger than 31.2 are marked with solid blue diamonds for EMD 1129 denoising data after 10 iterations. A helix backbone structure can be clearly identified. In {\bf b}, a helix function is parametrized based on fitting with marked positions in {\bf a}. In each circle, there are 12 independent blue color nodes, which represent the same type protein monomers  (we call them Type ``I" monomers). In {\bf c}, 12 black color nodes are added evenly on helix curve in each circle, representing Type ``II" protein monomers. Together with the blue nodes, they are used to compare with the EMD1129 experimental data, in which there are 24 proteins in each circle. In {\bf d}, three and four more layers of proteins are added to two ends of the helix curve to eliminate the boundary effect in our density function.  }
\label{fig:Fit_Curve}
\end{figure}

For cryo-EM data of low resolution or intermediate resolution, it is well-known that atomic scale models are unreliable. As such, coarse-grained models in terms of residues or even proteins can be useful. In this work, we propose a coarse-grained model for  microtubule.

The EMD-1129 data seriously suffer from noise as demonstrated in Fig.  \ref{fig:EMD1129} ({\bf a}). To build up a coarse-grained model, a topological denoising process as discussed in the previous section is employed. Surfaces extracted from denoising data are illustrated in Fig.  \ref{fig:EMD1129}. It can be seen that after ten iterations, the noise intensity is dramatically reduced and the basic geometry of the structure is preserved. More iterations are considered due to the requirement of persistent homology analysis, which will be discussed later in Section \ref{Sec:Mode_evaluation}.

Based on the denoising data processed with ten iterations, we analyze the structure of this microtubule intermediate and build up coarse-grained models.  Through the observation of different isosurfaces from the data it can be found that this microtubule intermediate has a unique helix backbone configuration. We assume that the center of each component protein has the largest electron density value. Through a threshold value of 31.2, we are able to identify centers of these component proteins and the helix backbone is thus constructed. Figure  \ref{fig:Fit_Curve} {\bf a} demonstrates the construction of our theoretical model with two types of protein monomers. A helix function as illustrated in Fig. \ref{fig:Fit_Curve}  {\bf b} is parametrized based on fitting with these marked positions. It is seen that in each circle, there are about 12 blue color nodes, representing 12 protein monomers of the same type which we denote as type ``I". However, there are  12  type ``II'' monomers missing in this model as they has a slightly lower electron density value. These type ``II" protein monomers are further accounted by adding 12 black  nodes evenly distributed on  the helix curve so that they can pair up with type ``I" protein monomers as demonstrated in  Fig. \ref{fig:Fit_Curve}  {\bf c}. Finally, to avoid the boundary effect in our density function evaluation, about three  layers are added to the top and bottom parts of the helix structure.

\begin{figure}
\begin{center}
\begin{tabular}{c}
\includegraphics[width=0.8\textwidth]{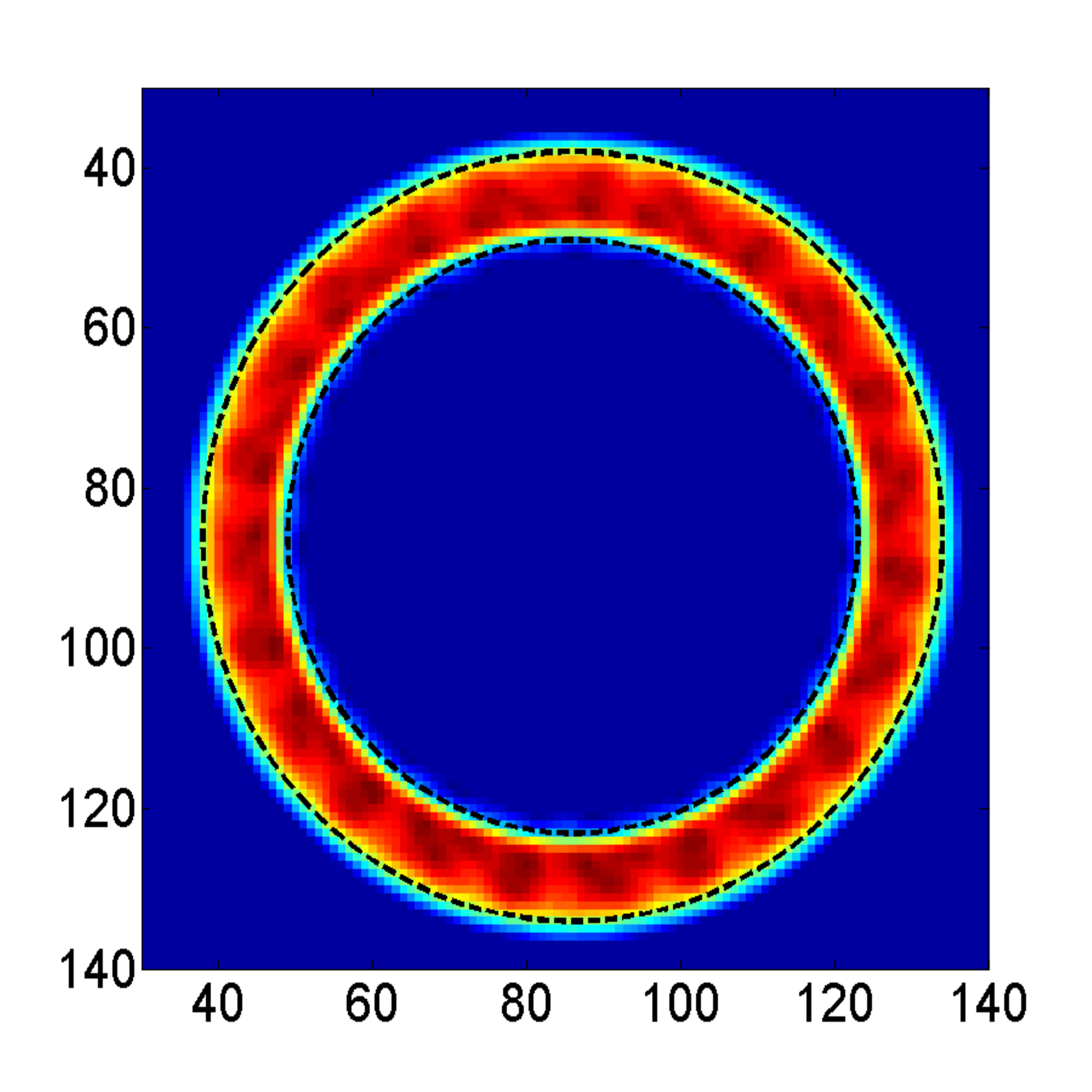}
\end{tabular}
\end{center}
\caption{Illustration of the fitting region. The average isovalues over the $z$-axis, i.e., $\frac{1}{n_z}\sum_k {\rho^e (x_i,y_j,z_k)}$ are given. The fitted region is indicated by two dash lines, i.e., the inner circle and the outer circle. Two circles share the same center point coordinate (86, 86) and their radius are 37 and 48 voxels, respectively. }
\label{fig:FiltrationMatrix}
\end{figure}

To avoid the complexities and present our persistent homology analysis directly and clearly, a simple coarse-grained model is considered. Basically, we use ellipsoids  to represent protein monomers. The density function of our microtubule intermediate model can be expressed as,
\begin{eqnarray}\label{ellipsoids}
\rho (x,y,z) = \sum_{i}W_i e^{-\left[ (\frac{x-x_i}{ (\sigma^x_i)})^2+ (\frac{y-y_i}{ (\sigma^y_i)})^2+ (\frac{z-z_i}{ (\sigma^z_i)})^2\right]},
\end{eqnarray}
where $\rho (x,y,z)$ is the density function of the model, parameter $W_i$ is the weight coefficient,  parameters $ (\sigma^x_i)$, $ (\sigma^y_i)$ and $ (\sigma^z_i)$ are ellipsoid radii. Coordinates $(x_i, y_i, z_i)$ denote the positions of protein monomer centers. To eliminate the boundary effect, the simulated models incorporate extra protein elements as illustrated in Fig.  \ref{fig:Fit_Curve} {\bf d}. All the above parameters are optimized by the least-square fitting using the denoising data.  It is found that in any $xy$-cross section, the electron density of microtubule tightly concentrates in a highly symmetric ring-band region as shown in Fig.  \ref{fig:FiltrationMatrix}. This ring-band region can be characterized by an inner circles and an outer circle. These circles share the same center at grid position (86, 86) and their radii are 37 and 48 voxels, respectively. As discussed in the literature \cite{Topf:2008}, only the regions that have sufficiently large density values  should be included in the fitting. In the present work, the fitting region is limited to the region within  two dash-line circles in the $xy$-cross section as illustrated in Fig.  \ref{fig:FiltrationMatrix}. In 3D, a thick-layer-cylinder region  that encompasses the structure is considered as the fitting region and is denoted as $V$.

To obtain the optimized fitting parameters,  two evaluation coefficients, i.e., cross-correlation coefficients (CCF)
\begin{eqnarray}
{\rm CCF}=\frac{\sum_{j\in V}{\rho_j^e} \sum_{j\in V}{\rho_j}}{\sqrt{\sum_{j\in V}{ (\rho_j^e)^2} \sum_{j\in V}{ (\rho_j)^2}}},
\end{eqnarray}
and correlation coefficients (CF)
\begin{eqnarray}
{\rm CF}=\frac{\sum_{j\in V}{ (\rho_j^e-\bar{\rho}_j^e)}\sum_{j\in V}{ (\rho_j-\bar{\rho}_j)}}{\sqrt{\sum_{j\in V}{ (\rho_j^e-\bar{\rho}_j^e)^2}\sum_{j\in V}{ (\rho_j-\bar{\rho}_j)^2}}}.
\end{eqnarray}
are used \cite{Topf:2008}. Here $\rho_i=\rho(x_j,y_j,z_j)$, $\rho_j^e$ is the experimental electron-density value at $(x_j,y_j,z_j)$  after 10  denoising iterations, $\bar{\rho}$ denotes the average of $\rho$ and parameter $V$ is the fitting region as stated above.
\begin{figure}
\begin{center}
\begin{tabular}{c}
\includegraphics[width=0.8\textwidth]{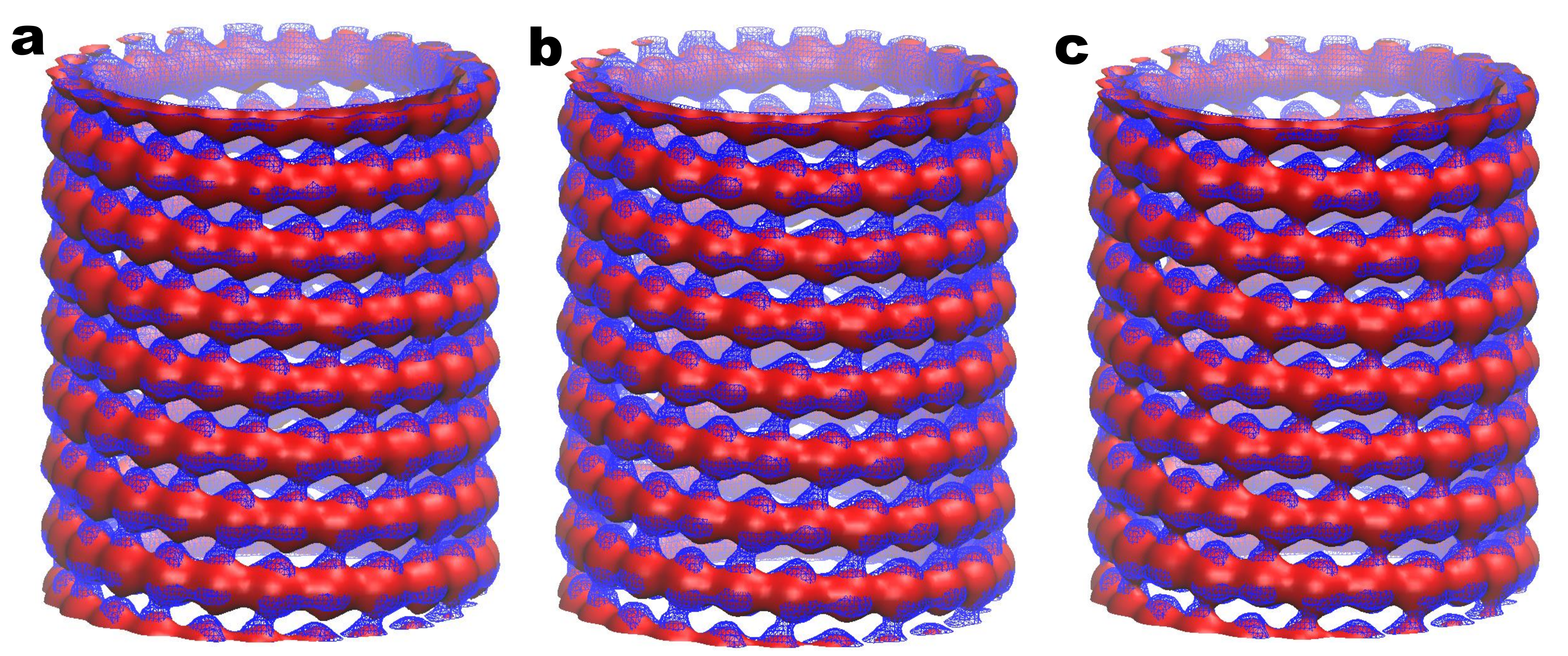}
\end{tabular}
\end{center}
\caption{Three theoretical models for microtubule structures constructed from  fitting the experimental data using proteins on helix backbone curve. Based on the coarse-grained representation, we use an ellipsoid to represent a protein monomer. In {\bf a}, only one type of ellipsoids is used. In {\bf b}, two types of ellipsoids with different weight functions are used. In {\bf c}, two types of ellipsoids with two different weight functions and modified locations are considered. The isosurfaces in {\bf a} and  {\bf b} look  similar to each other. However, they have dramatically different topological behaviors. In  {\bf c}, we systematically shift type ``II" monomers to form dimmers with type ``I" monomers. The blue meshed surfaces are obtained from the denoising data, and red solid surfaces are computed from the corresponding three theoretical models.}
\label{fig:Microtubule_Fitting}
\end{figure}

Based on the helix backbone configuration, three theoretical models are constructed for microtubule structure. Using the least square fitting, we determine the fitting parameters, i.e,  $W_i$ and  $\sigma$ in Eq. (\ref{ellipsoids}) for these models. Results are evaluated by aforementioned CCF and CF criteria.  In the first model, only one type of ellipsoids is used, i.e., one type of monomers  with $W_i=42$ for all  protein monomers. The second model has two types of monomers, see  Figure  \ref{fig:Fit_Curve}.  Two types of ellipsoids with $w_1=42$ and $w_2=38$ are considered by setting $\{W_i; W_i=w_1$ or $w_2\}$. The third model is dimer one. In this model, location modification is considered to generate dimers by shifting the type ``II" protein monomers simultaneously along the helix backbone closer to type ``I" protein monomers slightly. In this manner, we discriminate between  intra-dimer and inter-dimer distances. All ellipsoids are parametrized uniformly by assuming $ \sigma^x_i=24$, $ \sigma_i^y=24$ and $ \sigma_i^z=22$. The unit for these parameters is Angstrom (\AA), and voxel spacing is 4 \AA. The results are illustrated in Fig.  \ref{fig:Microtubule_Fitting}. All three models look similarly. Actually, the CCFs for the three models are 0.9601, 0.9607 and 0.9604, respectively. The CFs for them are 0.7392, 0.7436 and 0.7662, respectively. The difference in these coefficients is very small. Therefore, we cannot determine with a good confidence that one model is definitely better than  others. We therefore encounter a standard ill-posed inverse problem by using the structural optimization approach.

\subsection{Persistent homology based microtubule model evaluation}\label{Sec:Mode_evaluation}
\begin{figure}
\begin{center}
\begin{tabular}{c}
\includegraphics[width=0.9\textwidth]{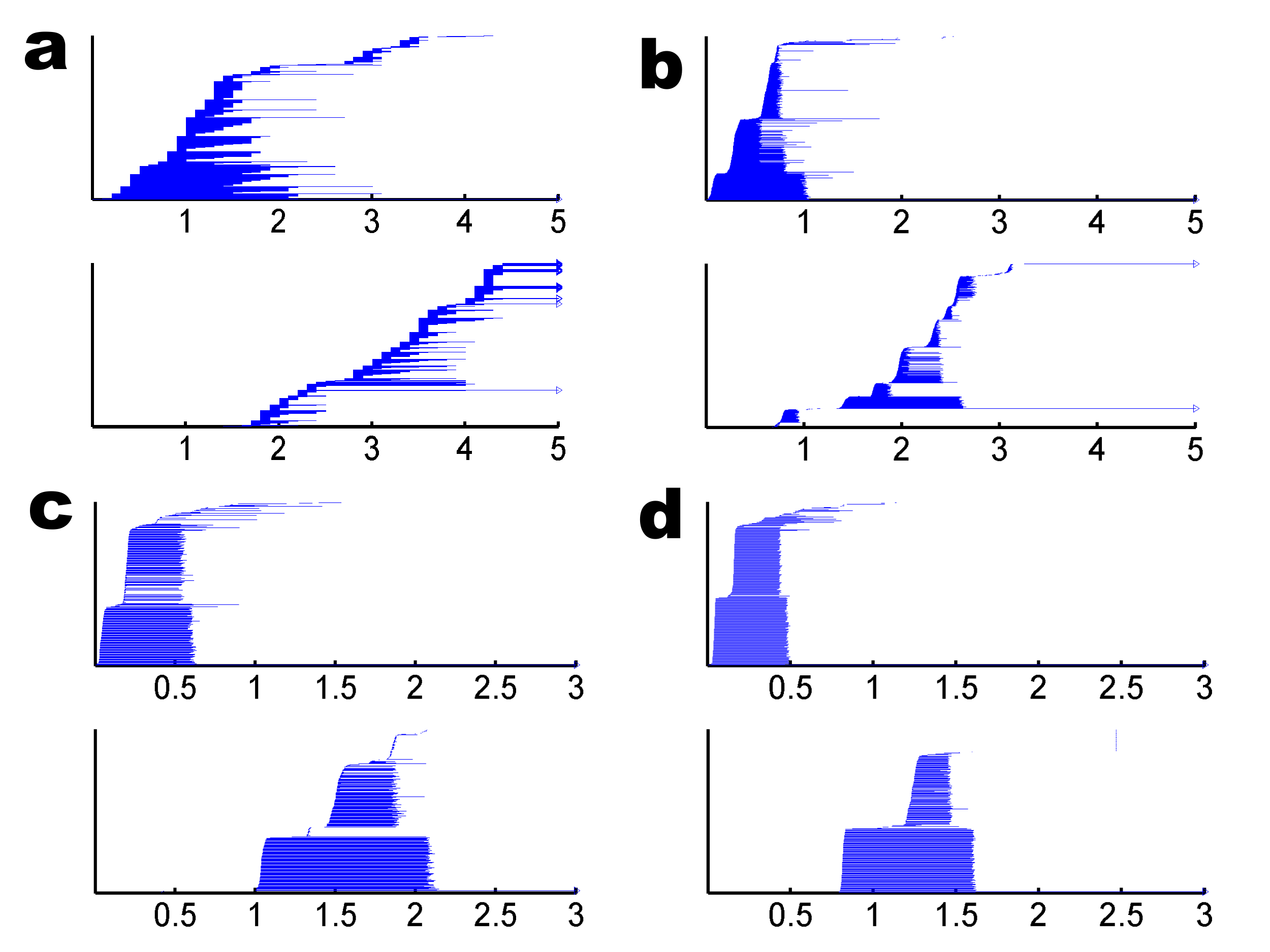}
\end{tabular}
\end{center}
\caption{ Topological persistence  of $\beta_0$ (top row) and $\beta_1$ (bottom row) generated from original and preprocessed EMD 1129 data. {\bf a} is the barcodes  of the original EMD1129 data. {\bf b}, {\bf c} and {\bf d} are barcodes for EMD-1129 data after 10, 20 and 40 denoising iterations. A special pattern, i.e., two individual bands of bars in both $\beta_0$ and $\beta_1$, persists in {\bf c} and {\bf d}.   }
\label{fig:1129_Barcodes_Polish}
\end{figure}

However, if we pay attention to the $\beta_1$ patterns of holes formed between the upper and lower helix circles as illustrated in Fig.  \ref{fig:Microtubule_Fitting}  ({\bf c}), it can been seen that only the third model preserves these features. These linkage properties are directly related to biomolecular flexibility and functional properties. Therefore it is important for us to characterize and capture them in our models. Unfortunately, least square optimization approach is insensitive to these little structural characteristics. Therefore, techniques that are sensitive to geometric variations are required to guide the structure determination. We show that persistent homology is a desirable technique for detecting geometric changes  in the rest of this section.

\begin{figure}
\begin{center}
\begin{tabular}{c}
\includegraphics[width=0.9\textwidth]{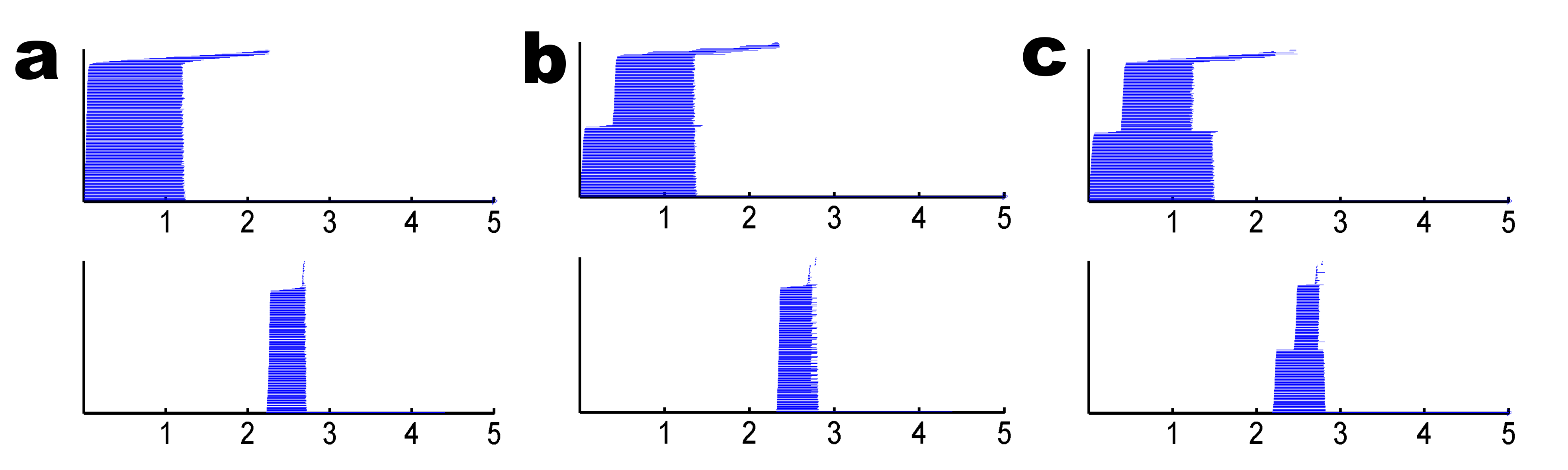}
\end{tabular}
\end{center}
\caption{Topological fingerprints of $\beta_0$ (top row) and $\beta_1$ (bottom row) generated from three theoretical  models as depicted in Fig.  \ref{fig:Microtubule_Fitting}. {\bf a} is the barcodes of the first model with only one type of monomers.  {\bf b} is the barcodes of the second fitted model with only two types of monomers using different weight functions. {\bf c} is barcodes of the third model with two types of monomers using different weight functions and modified locations. It can be seen that only the final model is able to capture the topological properties of the intrinsic topological fingerprints of cryo-EM data in Fig. \ref{fig:1129_Barcodes_Polish} {\bf d}. }
\label{fig:1129_Fit_Barcodes}
\end{figure}

To understand how the persistent homology can be employed to guide our model construction and evaluation, we investigate the topological  fingerprint  of the microtubule intermediate structure. As described earlier, due to the noise, a denoising process is required. The geometric flow based denoising algorithm is used with different numbers of iterations. The denoising data are carefully analyzed and the topological persistence  results are demonstrated in Fig.  \ref{fig:1129_Barcodes_Polish}. It can  be seen that a special pattern begin to emerge when the number of iterations approaches 20.  More specifically, groups of bars in both $\beta_0$ and $\beta_1$ panels appear with distinctive persisting patterns.

\begin{figure}
\begin{center}
\begin{tabular}{c}
\includegraphics[width=0.9\textwidth]{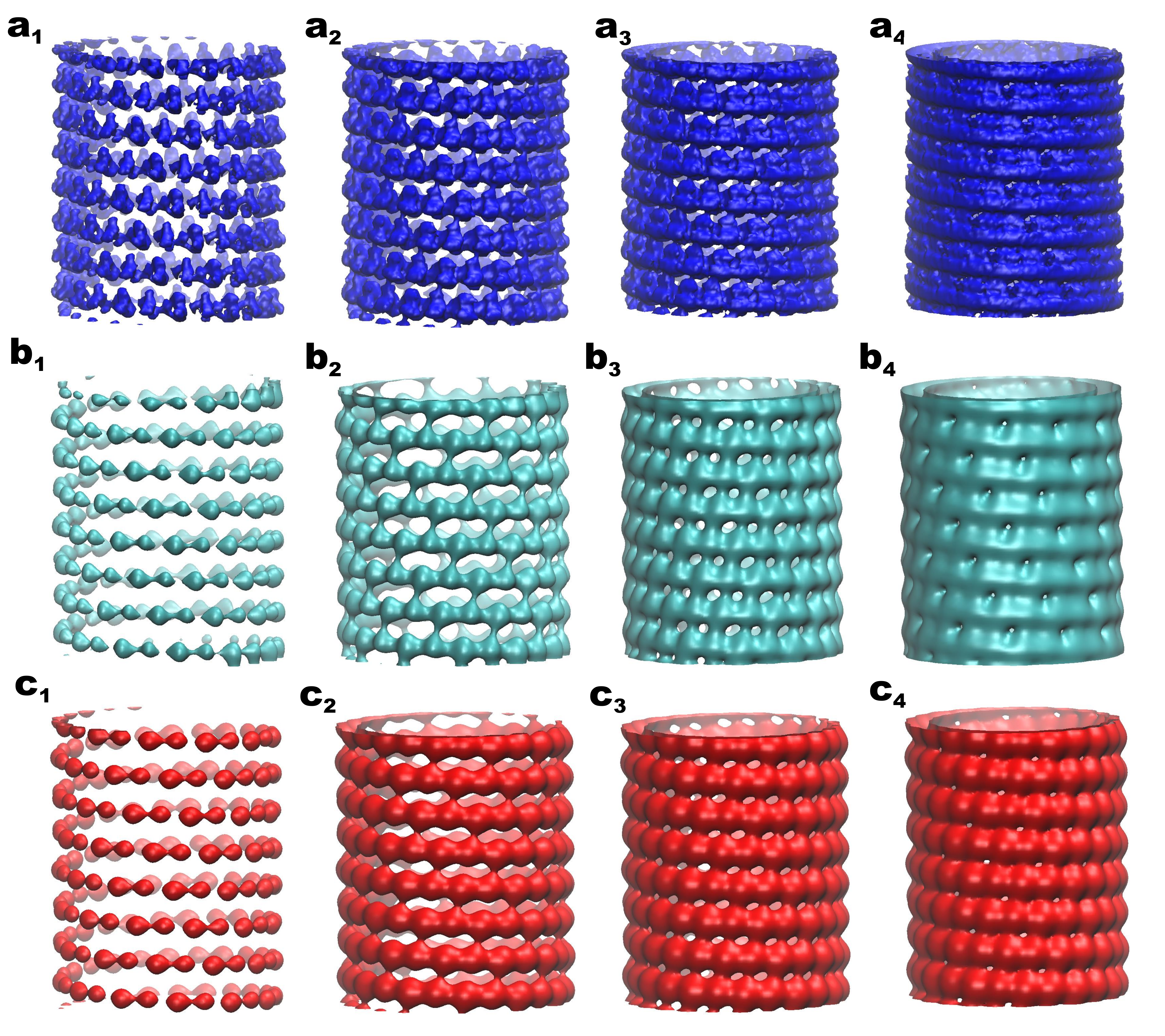}
\end{tabular}
\end{center}
\caption{The topological transitions of  microtubule geometry. Here  {\bf a},  {\bf b} and  {\bf c} are the original data, denoising data (40 iterations) and theoretical model (the third model), respectively. The subscripts 1 to 4 represent four topological transitions during the filtration process, i.e., hetero-dimmer formation, large circles formation, evolution of each large circles into two circles, and finally death of one of two circles. It can be seen that, these topological transitions are well-preserved in our denoising data and theoretical model, which explains the excellent topological consistency between them.}
\label{fig:1129_homology}
\end{figure}

In the $\beta_0$ panel of Fig.  \ref{fig:1129_Barcodes_Polish}, bars can be roughly groups into three parts from the top to the bottom, i.e., an irregular ``hair-like" part on the top, a narrow regular ``body" part in the middle and a large regular ``base" part in the bottom. Topologically, these parts represent different components in the microtubule intermediate structure. The irregular ``hair-like" part corresponds to the partial monomer structures located on the top and the bottom boundaries of the structure. As can be seen in Fig.  \ref{fig:Microtubule_Fitting}, each monomer has lost part of the structure at the boundary regions. The regular ``body" and ``base" parts are basically related to two types of monomers in the middle region where the  structure is free of boundary effect. From the barcodes, it can be seen that ``body" part has a later ``birth" time and earlier ``death" time compared with the ``base" barcode part. This is due to the reason that this type of monomers has relative lower electron density. As the filtration is defined to go from highest electron density values to lowest ones, their corresponding barcodes appear later. Their earlier death time, however, is due to the reason that they form dimers with the other type of monomers represented by the ``base" barcode part. It can be derived from these nonuniform behavior that monomers are not equally distributed along the helix backbone structure. Instead, two adjacent different types of monomers form a dimer first and then all these dimers  simultaneously connect with each other as the filtration goes on. Moreover, from the analysis in the previous section, it is obvious to see that the ``body" and ``base" parts are topological representations of type ``II" monomer and type ``I" monomer, respectively.

For the $\beta_1$ panel of Fig.  \ref{fig:1129_Barcodes_Polish}, there also exists a consistent pattern when the denoising process passing a certain stage. Two distinctive types of barcodes can be identified in the fingerprint, i.e., a shorter band of barcodes on the top and a longer band of bars on the bottom. Topologically, these $\beta_1$ bars correspond to the rings formed between two adjacent helix circles of monomers or dimers. During the filtration, dimers are formed between type ``I" and type ``II" monomers and soon after that, all dimers connect with each other and form the helix  backbone. As the filtration goes on, type ``I" monomers from the upper helix circle first connect with  type ``II" monomers at the lower circle. Geometrically, this means six monomers, three  (``I-II-I") from the upper layer and three (``II-I-II") from the lower layer,  form a circle. As the filtration goes further, this circle evolves into two circles when two middle monomers on two layers also connect. However, these two circles do not disappear simultaneously. Instead, one persists longer than the other.  This entire process generates the unique topological fingerprint in $\beta_1$  barcodes.

The topological fingerprint we extracted from the denoising process can be used to guide the construction and evaluation of our microtubule models. To this end,   we analyze the topological features of three theoretical models. Our persistent homology results for three models are demonstrated in Figs.  \ref{fig:1129_Fit_Barcodes} {\bf a}, {\bf b} and  {\bf c}, respectively. It can be seen that all the three models are able to capture the irregular ``hair" region in their $\beta_0$ barcode chart. From the topological point of view, the first model is the poorest one. It fails to capture the regular fingerprint patterns in both $\beta_0$ and $\beta_1$ panels of the original cryo-EM structure in   Fig.  \ref{fig:1129_Barcodes_Polish} {\bf d}. With two different weight functions to represent two types of monomers, the second model delivers a relatively better topological result. It is able to preserve part of the difference between type ``I" and type ``II" barcodes in the $\beta_0$ panel.  In $\beta_1$ panel, some nonuniform barcodes emerges. The persistent homology results are further improved in the third model when the intra-dimer and inter-dimer interactions are considered. In our third model,  fingerprint  patterns of the cryo-EM structure in both $\beta_0$ and $\beta_1$ panels of Fig.  \ref{fig:1129_Barcodes_Polish} {\bf d} are essentially  recovered by those of Fig.  \ref{fig:1129_Fit_Barcodes} {\bf c}. Even though their scales are different, their shapes are strikingly similar.

\subsection{Discussion}

The essential topological features that are associated with  major topological transitions of the original cryo-EM structure are illustrated in Figs.  \ref{fig:1129_homology} {\bf a$_1$},  {\bf a$_2$},  {\bf a$_3$}  and {\bf a$_4$}. As shown in  Figs.  \ref{fig:1129_homology}  {\bf b$_1$},  {\bf b$_2$},  {\bf b$_3$}  and {\bf b$_4$}, these features have  been well-preserved during the denoising process. Our best predicted model is depicted in  Figs.  \ref{fig:1129_homology}  {\bf c$_1$},  {\bf c$_2$},  {\bf c$_3$}  and {\bf c$_4$}. In these figure labels, subscripts $1,2,3 $ and $ 4$ denote four topological transition stages  in the filtration process, namely hetero-dimmer  formation, large circles formation, evolution of one large circle into two circles, and finally death of one of two circles. By the comparison of denoising results  (Figs. {\bf b$_1$, b$_2$, b$_3$} and {\bf a$_4$}) with original structures  (Figs. {\bf a$_1$, a$_2$, a$_3$} and {\bf a$_4$}), it is seen  that, in the noise reduction process, although some local geometric and topological details are removed, fundamental topological characteristics are well preserved. As illustrated in Fig.  \ref{fig:1129_Barcodes_Polish}, using the persistent homology description, these fundamental topological characteristics are well-preserved in topological persistence  patterns, which are further identified as fingerprints of the microtubule intermediate structure. We believe that  topological fingerprints are crucial  to the characterization, identification, and analysis  of the biological structure. As demonstrated in Figs.  \ref{fig:1129_homology} {\bf c$_1$},  {\bf c$_2$},  {\bf c$_3$}  and {\bf c$_4$}, once our model successfully reproduces the topological fingerprints, the simulated structure is able to capture the essential topological characteristics of the original one.  Moreover, through the analysis in Section \ref{Sec:Mode_evaluation}, it can be seen that to reproduce the topological fingerprint of  EMD 1129, two conditions are essential. The first is the creation  of two types of monomers. The second is the differentiation  of  intra-dimers and inter-dimers. Biologically, these requirements means: 1) there are two types of monomers, i.e., $\alpha$-tubulin monomers and $\beta$-tubulin monomers; and 2) intra-dimers and inter-dimers should behave differently from hetero-dimers.

It also should be noticed that a higher correlation coefficient may not guarantee the success of the model, especially when the original data is of  low resolution and low SNR. As can be seen in Section \ref{Sec:three_models}, our three theoretical models have very similar fitting coefficients. The second model even has a slightly  higher cross-correlation coefficients. However, only the third model is able to reproduce the essential topological features of the original cryo-EM data. This happens as  topological invariants, i.e., connected components, circles, loops, holes or void, tend to be very sensitive to  ``tiny" linkage parts, which  are almost negligible in the density  fitting process, compared to the major body part. We believe these linage parts play important roles in biological system especially in macroproteins and protein-protein complexes. Different linkage parts generate different connectivity, thus can directly influence biomolecular flexibility, rigidity,  and even its functions. By associates topological features with geometric measurements, our persistent homology analysis is able to distinguish  these connectivity parts. Therefore, persistent homology  is able to play a unique role in protein design, model evaluation and structure determination.

\section{Conclusion}

Cryo-electron microscopy (cryo-EM) is a major workhorse for the investigation  of subcellular structures, organelles and large multiprotein complexes. However, cryo-EM techniques and algorithms are far from mature, due to limited sample quality and or stability, low signal to noise (SNR), low resolution, and the high complexity of the underlying molecular structures. Persistent homology is a new branch of topology that is known for its potential in the characterization, identification and analysis (CIA) of big data.   In this work, persistent homology is, for the first time, employed for the cryo-EM data CIA.

Methods and algorithms for the geometric and topological modeling are presented. Here, geometric modeling, such the generation of density maps for proteins or other molecules,  is employed to create known data sets for validating  topological modeling algorithms. We demonstrate that cryo-EM density maps and  fullerene density data can be effectively analyzed by using  persistent homology.

Since topology is very sensitive to noise, the understanding of the topological signature of noise is a must in cryo-EM CIA. We first investigate the topological fingerprint of Gaussian noise. We reveal that for the Gaussian noise, its topological invariants, i.e.,  $\beta_0$, $\beta_1$ and $\beta_2$ numbers,  all exhibit the Gaussian distribution in the filtration space, i.e., the space of volumetric density isovalues.  At a low SNR,  signal and noise are inseparable in   the filtration space. However, after denoising with the geometric flow method,  there is clear separation between signal and noise for various topological invariants.  As such, a simple threshold can be prescribed to effectively remove noise. For the case of low SNR, the understanding of noise characteristic in the filtration space enable us to use persistent homology as an efficient means to monitor and control the noise removal process. This new strategy for noise reduction  is called topological denoising.

Persistent homology has been applied to the theoretical modeling of a  microtubule structure (EMD 1129). The backbone of the microtubule has a helix structure. Based on the helix structure, we propose  three theoretical models. The first model assumes that protein  monomers form the helix structure.  The second model adopts two types of protein monomers evenly distributed along helix chain. The last model utilizes a series of protein dimers along the helix chain. These models are fitted with experimental data by the least square optimization method. It is found that all the three models give rise to similar high correlation  coefficients with the experimental data, which indicates that the structural optimization is ill-posed. However, the topological fingerprints of three models are dramatically different. In the denoising process, the cryo-EM data of  the microtubule   structure demonstrate a consistent pattern which can be recognized as the intrinsic topological fingerprint of the microtubule   structure.  By careful examination of the fingerprint, we reveal two essential topological  characteristics which discriminate the protein dimers from the monomers. As such, we conclude that  only the third model, i.e., the protein dimer model,  is able to capture the intrinsic topological characteristics of the cryo-EM structure and must be the best model for the experimental data. It is believed that the present work offers a novel topology based strategy for resolving ill-posed inverse problems.

\section*{Acknowledgments}

This work was supported in part by NSF grants  DMS-1160352 and IIS-1302285,  NIH Grant R01GM-090208 and MSU  Center for Mathematical Molecular Biosciences initiative. The authors acknowledge the Mathematical Biosciences Institute for hosting valuable workshops.

\vspace{0.6cm}

\end{document}